\documentclass[onecolumn]{emulateapj}  
\usepackage{epstopdf}
\usepackage{ulem}
\usepackage{amssymb}
\usepackage{natbib}
\usepackage{times}
\usepackage{graphicx}
\usepackage[usenames,dvipsnames]{pstricks}
 \usepackage{psfrag}
\usepackage[colorlinks=true,linkcolor=blue,citecolor=blue]{hyperref}

\def\aj{AJ}
\def\araa{ARA\&A}
\def\apj{ApJ}
\def\apjl{ApJ}
\def\apjs{ApJS}
\def\apss{Ap\&SS}
\def\aap{A\&A}
\def\jcap{J. Cosmology Astropart. Phys.}
\def\mnras{MNRAS}
\def\na{New A}
\def\prd{Phys.~Rev.~D}
\def\sovast{Soviet~Ast.}
\def\nat{Nature}
\def\physrep{Phys.~Rep.}

\newcommand{\be}{\begin{equation}}
\newcommand{\ee}{\end{equation}}
\newcommand{\bary}{\begin{eqnarray}}
\newcommand{\eary}{\end{eqnarray}}
\newcommand{\en}{E_\nu}
\newcommand{\enc}{E_{\nu,c}}

\shorttitle{The radio galaxy IC310}
\shortauthors{Fraija et al.}
\begin{document}
\title{Modeling the Spectral Energy Distribution of the radio galaxy IC310}
\author{N. Fraija$^1$, A.~Marinelli$^2$, A.~Galv\'an-G\'amez$^1$ and E.~Aguilar-Ruiz$^1$}
  
\affil{$^1$Instituto de Astronom\' ia, Universidad Nacional Aut\'onoma de M\'exico, Circuito Exterior, C.U., A. Postal 70-264, 04510 M\'exico City, M\'exico\\
$^2$  I.N.F.N. \& Physics Institute Polo Fibonacci Largo B. Pontecorvo, 3 - 56127 Pisa, Italy}
\email{nifraija@astro.unam.mx, antonio.marinelli@pi.infn.it, agalvan@astro.unam.mx and eaguilar@astro.unam.mx}
%
%
%

\begin{abstract}
The radio galaxy IC310 located in the Perseus Cluster is one of the brightest objects in the radio and X-ray bands, and one of the  closest active galactic nuclei observed in very-high energies.   In GeV - TeV $\gamma$-rays, IC310 was detected in low and high flux states by the MAGIC telescopes from October 2009 to February 2010. Taking into account that the spectral energy distribution (SED) up to a few GeV seems to exhibit a double-peak feature and  that a single-zone synchrotron self-Compton (SSC) model can explain all of the multiwavelength emission except for the non-simultaneous MAGIC emission,   we interpret, in this work,  the multifrequency data set of the radio galaxy IC310 in the context of homogeneous hadronic and leptonic models.  In the leptonic framework, we present a multi-zone SSC model with two electron populations to explain the whole SED whereas for the hadronic model, we propose that a single-zone SSC model describes the SED up to a few GeVs and  neutral pion decay products resulting from p$\gamma$ interactions  could describe the  TeV - GeV $\gamma$-ray spectra. These interactions occur  when Fermi-accelerated protons interact with the seed photons around the SSC peaks.    We show that, in the leptonic model the minimum Lorentz factor of second electron population is exceedingly high $\gamma_e\sim10^5$ disfavoring this model, and in the hadronic model the required proton luminosity is not  extremely high $\sim 10^{44}$ erg/s, provided that charge neutrality between the number of electrons and protons is given.  Correlating the TeV $\gamma$-ray and neutrino spectra through photo-hadronic interactions, we find that the contribution of the emitting region of IC310 to the observed neutrino and ultra-high-energy cosmic ray fluxes are negligible. 

\end{abstract}


%
\section{Introduction}
Active galactic nuclei (AGN) are one of the most extreme astrophysical sources in the Universe.   These are compact, extremely luminous and often most variable shortest observed frequencies.  Perpendicular to the plane of the accretion material, magnetically driven and collimated relativistic plasma outflows are ejected away from the AGN \citep{2013MNRAS.430.2828L, 1987A&A...173..237L,2002ApJ...572..445L}.   They,  extensively studied from radio wavelengths to very-high-energy (VHE) $\gamma$-rays, are usually explained by the non-thermal  synchrotron self-Compton (SSC) model \citep{2010ApJ...719.1433A,2012ApJ...753...40F,1998ApJ...509..608T}, although hadronic models in some cases have been required \citep{2011ApJ...736..131A, 2002MNRAS.332..215A, 2001APh....15..121M,2003ApJ...586...79A}.\\
The radio galaxies, a type of AGN,  are elliptical galaxies with  two giant lobes  and misleading-bipolar jets of gas extending away from a central nucleus.  The radio galaxies  are generally described by the standard non-thermal  SSC model \citep{2010ApJ...719.1433A,2012ApJ...753...40F,1998ApJ...509..608T},  low-energy emission (from radio to optical) from synchrotron radiation and HE photons (from X-rays to $\gamma$-rays) from inverse Compton scattering emission.   This model with only one electron population,  predicts a spectral energy distribution (SED) that can be hardly extended up to higher energies than a few GeVs \citep{2005ApJ...634L..33G,2008A&A...478..111L}.  Therefore, some authors have claimed that the emission in the GeV - TeV energy range may have origin in different physical processes \citep{2011MNRAS.413.2785B,2005ApJ...634L..33G}.  The radio galaxy IC310, also called B0313+411 and J0316+4119, located in the Perseus Cluster at a redshift of z=0.00189 \citep{2002AJ....123.2990B} is one of brightest objects of the Perseus cluster of galaxies in the radio and X-ray bands.  Due to its kiloparsec-scale radio morphology, it has been classified as a head-tail or, more specifically, narrow-angle tail radio galaxy \citep{1998A&A...331..475F, 1998A&A...331..901S}. This radio galaxy being one of the fourth closest AGN observed in VHE $\gamma$-rays (after Cen A, M87 and NGC1275) is an excellent source for studying the physics of relativistic outflows.   It has been detected above 30 GeV by Large Area Telescope (LAT)-Fermi \citep{2010A&A...519L...6N}.  They reported the flux in two energy bands, $F_{30-100}=2.3^{+2.3}_{-1.2}\times 10^{-11}$ erg cm$^{-2}$ s$^{-1}$ between  30  and 100 GeV, and  $F_{100-300}=1.4^{+1.1}_{-0.6}\times 10^{-11}$ erg cm$^{-2}$ s$^{-1}$ from 100 to  300 GeV.   In addition, this source was detected by MAGIC telescopes at a high statistical significance of 7.6$\sigma$ in 20.6 hr of stereo data.  The TeV $\gamma$-ray flux of $(3.1\pm 0.5)\times 10^{-12}$ cm$^{-2}$ s$^{-1}$ (between 2009 October and 2010 February) was flat with a differential spectral index of 2.00$\pm$0.14 \citep{2010ApJ...723L.207A}.   The indications for flux variability on times scales of months and years were reported.  The variability time scales of years was confirmed by the non-detection of the source reported in 2010 August - 2011 February \citep{2012A&A...539L...2A}.  A re-analysis of the MAGIC stereo-observation taken during the same period was performed \citep{2014A&A...563A..91A}.  After fitting the spectra at high and low state  between  0.12  and  8.1 TeV with a simple power law, the observed fluxes reported by authors were $(4.28\pm0.21\pm0.73)\times 10^{-12}$ TeV$^{-1}$ cm$^{-2}$ s$^{-1}$ and $(0.608\pm0.037\pm0.11)\times 10^{-12}$ TeV$^{-1}$ cm$^{-2}$ s$^{-1}$, respectively \citep{2014A&A...563A..91A}. \\
%
%
Located at the South Pole,  the IceCube telescope was designed to detect the interactions of neutrinos. With a volume of one cubic kilometer of ice and almost four years of data taking, the IceCube telescope reported with the High-Energy Starting Events (HESE)\footnote{http://icecube.wisc.edu/science/data/HE-nu-2010-2014} catalog a sample of 54 extraterrestrial neutrino events in the TeV - PeV energy range.   Arrival directions of these events are compatible with an isotropic distribution and possible extragalactic origin \citep{2014PhRvD..90b3010A, 2015ApJ...815L..25G}.  Hadronic processes producing a fraction of the observed neutrino events revealed by the IceCube  via the acceleration of comic rays in radio galaxies have been explored \citep{2011A&A...531A..30R,2016MNRAS.455..838K, 2012PhRvD..85d3012S, 2014ApJ...783...44F, 2014MNRAS.441.1209F,2014arXiv1410.8549M}.\\
In this work we introduce leptonic and hadronic models to describe the broadband SED observed in the radio galaxy IC310.  In the leptonic framework, we present a multi-zone SSC model with two electron populations  to explain the whole SED whereas for the hadronic model, we propose that a single-zone SSC model describes the SED up to a few GeVs and  neutral pion decay products the proton-photon (p$\gamma$) interactions could model the $\gamma$-ray spectra (high and low state) at the GeV - TeV energy range.   Correlating the TeV $\gamma$-ray and neutrino spectra through photo-hadronic interactions, we estimate the neutrino and  ultra-high-energy cosmic ray (UHECR) fluxes.
\section{Theoretical Model}
Injected electrons and protons are accelerated in the emitting region  which moves at relativistic velocities.  These particles confined at the emitting region by magnetic fields are expected to cool down by synchrotron radiation, Compton scattering emission and photo-hadronic processes.  We hereafter use primes (unprimes) to define the quantities in a comoving (observer) frame, the universal constants   c=$\hbar$=1 in natural units and the values of cosmological parameters $H_0=$ 71 km s$^{-1}$ Mpc$^{-1}$, $\Omega_m=0.27$, $\Omega_\Lambda=0.73$  \citep{2003ApJS..148..175S}. 
\subsection{Synchrotron radiation}
\cite{1962SvA.....6..167K} showed that if the spectrum of injected electrons is {\small $\propto \gamma_e^{-\alpha_e}$}, then the spectrum varies with time as a result of losses due to emission in the magnetic field.  This variation differs for different energy intervals: {\small $N_e(\gamma_e)   = N_{0,e} \gamma_e^{-\alpha_e}$} for {\small $\gamma_{\rm e,min}<\gamma_e < \gamma_{\rm e,c}$} and  {\small $N_{0,e}\gamma_{e,c} \gamma_e^{-(\alpha_e+1)}$} for {\small $\gamma_{\rm e,c} \leq  \gamma_e<\gamma_{\rm e,max}$}, where  $N_{0,e}$ is the proportionality electron constant, $\alpha_e$ is the power index of the electron distribution and $\gamma_{e,i}$ are  the electron Lorentz  factors. The index $i$ is min, c or max for minimum, cooling and maximum, respectively.   Assuming that a fraction of total  energy density is given to accelerate electrons  $U_e=m_e \int\gamma_e N_e(\gamma_e)d\gamma_e$ ($m_e$ is the electron mass),  the electron luminosity and the minimum electron Lorentz factor are {\small $L_e=4\pi\, \delta^2_D\,r^2_d\,U_e$} and {\small $\gamma_{\rm e,min}=\frac{(\alpha_e-2)}{m_e(\alpha_e-1)}\,\frac{U_e}{N_e}$}, respectively. Here $r_d$ is the size of the emitting region and $\delta_D$ is Doppler factor.   The electron distribution submerged in a magnetic field $B=\sqrt{8\pi\,U_B}$ cools down following the cooling synchrotron time scale $t'_c=\frac{3m_e}{4\sigma_T}\,\,U^{-1}_B\,\gamma^{-1}_e$, with  $\sigma_T=6.65 \times 10^{-25}\, {\rm cm^2}$ the Thomson cross-section.  Comparing the synchrotron time scale with the dynamic scale $t'_d\simeq r_d/\delta_D$, we get that the cooling Lorentz factor is {\small$\gamma_{\rm e,c}= \frac{3\,m_e}{4\,\sigma_T}\,(1+Y)^{-1}\, \delta_D\, U_B^{-1}\,r_d^{-1}$}.  Here the Compton parameter is {\small $Y=\frac{\eta\,U_e}{U_B}$ for $\frac{\eta\,U_e}{U_B}\ll 1$ and {\small$\left(\frac{\eta\,U_e}{U_B}\right)^{1/2}$ for  $\frac{\eta\,U_e}{U_B}\gg 1$, with $\eta=(\gamma_{\rm e,c}/\gamma_{\rm e,min})^{2-\alpha_e}$ given for slow cooling and $\eta=1$ for fast cooling \citep{2001ApJ...548..787S}.    By considering that acceleration $t'_{acc}\simeq\sqrt\frac\pi2m_e/q_e\, U^{-1/2}_B\,\gamma_e$ and cooling time scales are similar, it is possible to write the maximum electron Lorentz factor as {\small $ \gamma_{\rm e,max}=\left(\frac{9\, q_e^2}{8\pi\,\sigma_T^2}\right)^{1/4}\,U_B^{-1/4}$ where $q_e$ is the electric charge. It is worth noting that the direction of the average magnetic field at the shock is shown to have a large effect.  If the perpendicular diffusion coefficient is much smaller than the parallel coefficient, particles can gain much more energy if the shock is quasi-perpendicular than quasi-parallel. In this case, the acceleration time scale  could be even shorter \citep{1987ApJ...313..842J}.    Taking into account the synchrotron emission ${\small \epsilon_\gamma(\gamma_{e,i})=\sqrt{\frac{8\pi q_e^2}{m_e^2}}\,\delta_D\, U_B^{1/2}\, \gamma^2_{e,i}}$ and from the electron Lorentz factors,  the synchrotron break energies are
\begin{eqnarray}\label{synrad}
\epsilon^{\rm syn}_{\rm \gamma,m} &=& \frac{\sqrt{8\pi}\,q_e}{m_e}\,\delta_D\,U_B^{1/2}\,\gamma^2_{\rm e,min}\,,\cr
\epsilon^{\rm syn}_{\rm \gamma,c} &=&\frac{ 9\sqrt{2\pi}\,q_e\,m_e}{8\,\sigma_T^2}\, (1+Y)^{-2}\,\delta_D^3\, U_B^{-3/2}\, r_d^{-2}\,,\cr
\epsilon^{\rm syn}_{\rm \gamma, max} &=&\frac{3\,q_e^2}{m_e\,\sigma_T}\, \delta_D\,.
\end{eqnarray}
The observed synchrotron spectrum can be written as \citep{1994hea2.book.....L, 1986rpa..book.....R}
{\small
\begin{eqnarray}
\label{espsyn}
\left[\epsilon^2_\gamma N(\epsilon_\gamma)\right]_{\rm \gamma, syn}= A_{\rm \gamma,syn}\cases{ 
(\frac{\epsilon_\gamma}{\epsilon^{\rm syn}_{\rm \gamma,m}})^\frac43    &  $\epsilon_\gamma < \epsilon^{\rm syn}_{\rm \gamma,m}$,\cr
 (\frac{\epsilon_\gamma}{\epsilon^{\rm syn}_{\rm \gamma,m}})^{-\frac{\alpha_e-3}{2}}  &  $\epsilon^{\rm syn}_{\rm \gamma,m} < \epsilon_\gamma < \epsilon^{\rm syn}_{\rm \gamma,c}$,\cr
(\frac{\epsilon^{\rm syn}_{\rm \gamma,c}}{\epsilon^{\rm syn}_{\rm \gamma,m}})^{-\frac{\alpha_e-3}{2}}    (\frac{\epsilon_\gamma}{\epsilon_{\rm \gamma,c}})^{-\frac{\alpha_e-2}{2}},           &  $\epsilon^{\rm syn}_{\rm \gamma,c} < \epsilon_\gamma < \epsilon^{\rm syn}_{\rm \gamma,max}$\,,\cr
}
\end{eqnarray}
}
\noindent where 
\be\label{Ae}
A_{\rm \gamma,syn}=\frac{4\,\sigma_T}{9}\, d^2_z\,\delta^3_D\,U_B\,r_d^3\,N_e\,\gamma^2_{\rm e,min}\,,
\ee
is the proportionality constant of synchrotron and $d_z$ is the distance between IC310 and Earth. 
\subsection{Compton scattering emission}
Fermi-accelerated electrons in the emitting region can upscatter synchrotron photons up to higher energies as {\small $\epsilon^{\rm ssc}_{\gamma,{\rm (m,c,max)}}\simeq  \gamma^2_{e,{\rm (min,c,max)}} \epsilon^{\rm syn}_{\gamma,{\rm (m,c, max)}}$}.    From  the electron Lorentz factors and the synchrotron break energies (eq. \ref{synrad}), we get that the Compton scattering break energies are given in the form  
\begin{eqnarray}\label{icrad}
\epsilon^{\rm ssc}_{\rm \gamma,m} &=& \frac{\sqrt{8\pi}\,q_e}{m_e}\,\delta_D\,U_B^{1/2}\,\gamma^4_{\rm e,min},\cr
\epsilon^{\rm ssc}_{\rm \gamma,c} &=&\frac{ 81\sqrt{2\pi}\,q_e\,m_e^3}{128\,\sigma_T^4}\,(1+Y)^{-4}\,\delta_D^5\, U_B^{-7/2}\, r_d^{-4},\cr
\epsilon^{\rm ssc}_{\rm \gamma, max} &=&\frac{9\,q_e^3}{2\sqrt{2\pi}\,m_e\,\sigma^2_T}\, \delta_D\,U_B^{-1/2}\,.
\end{eqnarray}
The Compton scattering spectrum  obtained  as a function of the synchrotron spectrum (eq. \ref{espsyn}) is
{\small
\begin{eqnarray} \label{espic}
\left[\epsilon^2_\gamma N(\epsilon_\gamma)\right]_{\rm \gamma, ssc}=A_{\rm \gamma,ssc} \cases{ 
(\frac{\epsilon_\gamma}{\epsilon^{\rm ssc}_{\rm \gamma,m}})^\frac43    &  $\epsilon_\gamma < \epsilon^{ssc}_{\gamma,m}$,\cr
 (\frac{\epsilon_\gamma}{\epsilon^{\rm ssc}_{\rm \gamma,m}})^{-\frac{\alpha_e-3}{2}}  &  $\epsilon^{\rm ssc}_{\rm \gamma,m} < \epsilon_\gamma < \epsilon^{\rm ssc}_{\rm \gamma,c}$,\cr
(\frac{\epsilon^{\rm ssc}_{\rm \gamma,c}}{\epsilon^{ssc}_{\gamma,m}})^{-\frac{\alpha_e-3}{2}}    (\frac{\epsilon_\gamma}{\epsilon^{ssc}_{\gamma,c}})^{-\frac{\alpha_e-2}{2}}          &  $\epsilon^{\rm ssc}_{\rm \gamma,c} < \epsilon_\gamma < \epsilon^{\rm ssc}_{\rm \gamma,max} $\,,
}\cr
\end{eqnarray}
}
where $A_{\rm \gamma,ssc}=Y\,[\epsilon^2_\gamma N_\gamma(\epsilon_\gamma)]_{\rm max}^{\rm syn}$ is the proportionality constant of inverse Compton scattering spectrum.
\subsection{Photo-hadronic Interactions}
Radio galaxies have been proposed as powerful accelerators of charged particles through the Fermi acceleration mechanism \citep{2009NJPh...11f5016D},  magnetic reconnection \citep{2015arXiv150407592K,2005Ap&SS.298..115L} and the magnetized accretion disk working as an electric dynamo \citep{1976Natur.262..649L, 1987ApJ...315..504L,1994ApJ...437..136L}.     We consider a proton population described as  a simple power law given by
\be\label{prot_esp}
\left(\frac{dN}{dE}\right)_p=A_p\,E_p^{-\alpha_p}\,,
\ee
with $A_p$ the proportionality constant and $\alpha_p$ the spectral power index of the proton population.  From eq. (\ref{prot_esp}), the proton density can be written as  {\small $U_p=\frac{L_p}{4\pi\,\delta^2_D\,r^2_d}$}   with the proton luminosity given by {\small $L_p= 4\,\pi\,d^2_z A_p\,\int \,E_p\,E_p^{-\alpha_p}  dE_p$}.    Fermi-accelerated protons  lose their energies by electromagnetic channels and hadronic interactions.  Electromagnetic channels such as proton synchrotron radiation and inverse Compton will not be considered here, we will only assume that protons cool down  by p$\gamma$ interactions at the emitting region of the inner jet. Charged ($\pi^+$) and neutral ($\pi^0$) pions are obtained from p$\gamma$ interaction with a fraction of 2/3 and 1/3 for {\small $p\,\pi^{0}$} and {\small $n\pi^{+} $}, respectively.
After that neutral pion decays into photons, $\pi^0\rightarrow \gamma\gamma$,  carrying $20\% (\xi_{\pi^0}=0.2)$ of the proton's energy $E_p$.   The efficiency of the photo-pion production is \citep{1968PhRvL..21.1016S, PhysRevLett.78.2292}
{\small
\begin{equation}\label{eficiency}
f_{\pi^0} \simeq \frac {t_{\rm dyn}} {t_{\pi^0}}  =\frac{r_d}{2\gamma^2_p}\int\,d\epsilon_\gamma\,\sigma_\pi(\epsilon_\gamma)\,\xi_{\pi^0}\,\epsilon_\gamma\int dx\, x^{-2}\, \frac{dn_\gamma}{d\epsilon_\gamma} (\epsilon_\gamma=x)\,,
\end{equation}
}
where $dn_\gamma/d\epsilon_\gamma$ is the spectrum of seed photons,  $\sigma_\pi(\epsilon_\gamma)$ is the cross section of pion production and $\gamma_p$ is the proton Lorentz factor. Solving the integrals we obtain 
{\small
\bary
f_{\pi^0} \simeq \frac{\sigma_{\rm p\gamma}\,\Delta\epsilon_{\rm res}\,\xi_{\pi^0}\, L_{\rm \gamma,IC}}{4\pi\,\delta_D^2\,r_d\,\epsilon_{\rm pk,ic}\,\epsilon_{\rm res}}
\cases{
\left(\frac{\epsilon^{\pi^0}_{\gamma,c}}{\epsilon_{0}}\right)^{-1} \left(\frac{\epsilon^{\pi^0}_{\gamma}}{\epsilon_{0}}\right)       &  $\epsilon_{\gamma} < \epsilon^{\pi^0}_{\gamma,c}$\cr
1                                                                                                                                                                                                            &   $\epsilon^{\pi^0}_{\gamma,c} < \epsilon_{\gamma}$\,,\cr
}
\eary
}
where $\Delta\epsilon_{\rm res}$=0.2 GeV,  $\epsilon_{\rm res}\simeq$ 0.3 GeV, $\epsilon_{\rm pk,ic}$ is the energy of the second SSC peak, $\epsilon_0$ is the normalization energy and  $\epsilon^{\pi^0}_{\gamma,c}$ is the break photon-pion energy is  {\small $\epsilon^{\pi^0}_{\gamma,c}\simeq 31.87\,{\rm GeV}\, \delta_D^2\, \left(\frac{\epsilon_{\rm pk,ic}}{ {\rm MeV}}\right)^{-1}$}.
It is worth noting that the target photon density is {\small $n_{\gamma}\simeq\frac{L_{\rm \gamma,IC}}{4\pi r^2_d\,\epsilon_{\rm pk,ic}}$} and the optical depth is {\small $\tau_{\gamma}\simeq\frac{L_{\rm \gamma,IC}\,\sigma_T}{4\pi r_d\,\delta_D\epsilon_{\rm pk,ic}}$}\,.  
Taking into account that  photons released  in the range $\epsilon_\gamma$ to $\epsilon_\gamma + d\epsilon_\gamma$ by protons in the range   $E_p$ and $E_p + dE_p$ are $f_{\pi^0}\,E_p\,(dN/dE)_p\,dE_p=\epsilon_{\pi^0,\gamma}\,(dN/d\epsilon)_{\pi^0,\gamma}\,d\epsilon_{\pi^0,\gamma}$, then photo-pion spectrum is given by
{\small
\bary
\label{pgammam}
\left[\epsilon^2_\gamma N(\epsilon_\gamma)\right]_{\rm \gamma, \pi^0}= A_{\rm p\gamma}  \cases{
\left(\frac{\epsilon^{\pi^0}_{\gamma,c}}{\epsilon_{0}}\right)^{-1} \left(\frac{\epsilon_{\gamma}}{\epsilon_{0}}\right)^{-\alpha_p+3}          &  $ \epsilon_{\gamma} < \epsilon^{\pi^0}_{\gamma,c}$\cr
\left(\frac{\epsilon_{\gamma}}{\epsilon_{0}}\right)^{-\alpha_p+2}                                                                                        &   $\epsilon^{\pi^0}_{\gamma,c} < \epsilon_{\gamma}$\,,\cr
}
\eary
}
\noindent where the proportionality constant  $A_{\rm p\gamma}$  is in the form
\be\label{Apg}
A_{\rm p\gamma}= \frac{L_{\rm \gamma,IC}\,\sigma_{\rm p\gamma}\,\Delta\epsilon_{\rm res}\,\epsilon^2_0\,\left(\frac{2}{\xi_{\pi^0}}\right)^{1-\alpha_p}}{4\pi\,\delta_D^2\,r_d\,\epsilon_{\rm pk,ic}\,\epsilon_{\rm res}}\,A_p\,.
\ee
The previous equations indicate that the value of $A_p$ is normalized using the TeV $\gamma$-ray flux.
\section{High energy neutrino expectation}
Photo-hadronic interactions in the emitting region (see subsection 3.2) also generate neutrinos through the charged pion decay products ($\pi^{\pm}\rightarrow \mu^\pm+  \nu_{\mu}/\bar{\nu}_{\mu} \rightarrow  e^{\pm}+\nu_{\mu}/\bar{\nu}_{\mu}+\bar{\nu}_{\mu}/\nu_{\mu}+\nu_{e}/\bar{\nu}_{e}$).  Taking into account the distance of IC310, the neutrino flux ratio created on the source (1 : 2 : 0 ) will arrive on the standard ratio \citep[1 : 1 : 1; e.g. see][]{2008PhR...458..173B}.  The neutrino spectrum produced by the photo-hadronic interactions is
{\small
\bary
\label{pgammam}\label{espneu1}
\left[\en^2 N(\epsilon_\nu)\right]_\nu= A_\nu \epsilon^2_0 \cases{
\left(\frac{\en}{\epsilon_0 }\right)^2                            &  $ \en < \enc$\cr
\left(\frac{\en}{\epsilon_0 }\right)^{2-\alpha_{\nu}}     &   $\enc < \en$\,,\cr
}
\eary
}
with the proportionality factor given by, {\small $A_{\nu}=A_{\rm p\gamma}\,\epsilon_0^{-2}\, 2^{-\alpha_p}$} \citep[see] [and reference therein]{2007Ap&SS.309..407H}.  The neutrino flux is detected when it interacts inside the instrumented volume. Considering the probability of interaction for a neutrino with energy $E_\nu$ in an effective volume with density ($\rho_{ice}$), the number of expected neutrino events after a period of time $T$ is  
{\small
\be
N_{\rm ev} \approx\,T \rho_{\rm ice}\,N_A\,\epsilon_0\int_{E_{\rm \nu,th}} V_{\rm eff}(E_\nu) \sigma_{\rm \nu N}(E_\nu) \,A_\nu\left(\frac{E_{\nu}}{\epsilon_0}\right)^{-\alpha_p}\, dE_\nu\,,
\label{numneu1}
\ee
}
where $N_A$ is the Avogadro number, $\sigma_{\nu N}(E_\nu)$ is the the charged current cross section, $E_{\nu,th}$ is the  energy threshold and $V_{eff}$ is the effective volume of IceCube.\\
\section{Ultra-high-energy cosmic rays}
Radio galaxies have been proposed as potential sources where  particles  could be accelerated up to UHEs \citep{1963SvA.....7..357G,1963SvA.....6..465S}.   We consider that the proton spectrum given by a simple power law is spread out as high as can be confined in the emitting region.
\subsection{Hillas Condition}
Requiring that super massive black holes (BHs) have the power to accelerate particles  up to UHEs through Fermi processes,  protons accelerated in the emitting region are confined by the Hillas condition \citep{1984ARA&A..22..425H}. Although this requirement is a necessary condition and acceleration of UHECRs in AGN jets \citep{2012ApJ...749...63M,2012ApJ...745..196R,2010ApJ...719..459J},  it is far from trivial (see e.g., \cite{2009JCAP...11..009L} for a more detailed energetic limits). The Hillas criterion is defined by
\be\label{Emax}
E_{\rm p,max}\simeq \frac{Zq_e}{\phi}\,B\,r_d\,\Gamma\,,
\ee
where $Z$ is the atomic number, $\phi\simeq$ 1 is the acceleration efficiency factor and $\Gamma=\frac{1\pm\sqrt{1-(1-\cos^2\theta)(1+\delta^2_D\cos^2\theta)}}{\delta_D(1-\cos^2\theta)}$ is the bulk Lorentz factor with $\theta$ is the viewing angle.    Similarly, supposing that the BH jet has the power also to accelerate particles  up to UHEs through Fermi processes, then during flaring intervals for which the apparent isotropic luminosity can reach $\approx 10^{45}$ erg s$^{-1}$ and  from the equipartition magnetic field $\epsilon_B$ the maximum particle energy of accelerated UHECRs can achieve values as high as  \citep{2009NJPh...11f5016D}
\be
E_{\rm max}\approx 10^{20}\,\frac{Zq_e\,\epsilon^{1/2}_B}{\phi\, \Gamma}\,\left(\frac{L_p}{10^{45}\,{\rm erg/s} }\right)^{1/2}\, {\rm eV}
\ee
\subsection{Deflections}
UHECRs traveling from source to Earth are randomly deviated by galactic (B$_G$) and extragalactic (B$_{EG}$) magnetic fields which play important roles on cosmic rays.  Requiring that magnetic fields are homogeneous and quasi-constant,  the deflection angle due to the B$_G$ is {\small $\psi_{\rm G}\simeq 4^{\circ}\left(\frac{57 EeV} {E_{p,th}}\right) \int^{L_G}_0  | \frac{dl}{{\rm kpc}}\times \frac{B_G}{4\,{\rm \mu G}} |$} and due to B$_{EG}$ is {\small $\psi_{\rm EG}\simeq 4^{\circ}\left(\frac{57 EeV} {E_{p,th}}\right) \,\left(\frac{L_{\rm EG}}{100\, {\rm Mpc}}\right)^{\frac12}\,\left(\frac{l_c}{1\, {\rm Mpc}}\right)^{\frac12}\,\left( \frac{B_{\rm EG}}{1\,{\rm nG}} \right)$} \citep{1997ApJ...479..290S},
where L$_{\rm G}$ corresponds to the distance of our Galaxy (20 kpc), $l_c$ is the coherence length and $E_{\rm p,th}$ is the threshold proton energy.   Due to the strength of extragalactic ($B_{\rm EG}\simeq$ 1 nG) and galactic ($B_{\rm G}\simeq$ 4 $\mu$G) magnetic fields,   UHECRs are deflected; firstly, $\psi_{\rm EG}\simeq 4^{\circ}$ and after $\psi_G\simeq 4^{\circ}$, between the true direction to the source and the observed arrival direction, respectively.  Estimation of the deflection angles could associate the transient UHECR sources with the HE neutrino and $\gamma$-ray fluxes. 
\subsection{UHECR Flux}
\paragraph{Telescope Array experiment}.  With an area of $\sim$ 700 km$^2$,  Telescope Array (TA) experiment at  Millard Country (Utah) study UHECRs with energies above 57 EeV (for details see \citealp{2012NIMPA.689...87A}).   The TA  exposure is $\frac{\Xi\,t_{op}\, \omega(\delta_s)}{\Omega}$=$\frac{(5)\,7}{\pi}\times10^2\,\rm km^2\,yr$, with  $\omega(\delta_s)$ the exposure correction factor \citep{2001APh....14..271S}. 
\paragraph{Pierre Auger observatory}.  Designed to determine the arrival directions and energies of cosmic rays above  57 EeV, the Pierre Auger observatory (PAO) is made by four fluorescence telescope arrays and 1600 surface detectors (for details see \citealp{2008APh....29..188P}).  The exposure of this observatory is $\frac{9\times10^3\,\omega(\delta_s)}{\pi}\,\rm km^2\,yr$ \citep{2007Sci...318..938P, 2008APh....29..188P}.  
\paragraph{The High Resolution Fly's eye observatory}.  The HiRes observatory with an exposure of (3.2 - 3.4) $\times 10^3$ km$^2$ year sr measured the flux of UHECRs using the stereoscopic air fluorescence technique  from 1997 to 2006   (for details see \citealp{2005ApJ...622..910A,2009APh....32...53H}).\\
The expected number and the flux of UHECRs above an energy $E_{p,th}$ are given by
\bary
N_{\rm \tiny UHECR}=\frac{\Xi\,t_{\rm op}\, \omega(\delta_s)}{(\alpha_p-1)\Omega}\,A_p  \int_{E_{\rm p,th}} E_{p}^{-\alpha_p} dE_p\,,
\label{nUHE1}
\eary
and 
\be
F_{\rm \tiny UHECR}=A_p\,\int_{E_{\rm p,th}} \,E_p\,E_p^{-\alpha_p}  dE_p\,,
\ee
respectively,  where the value of $A_p$ is normalized with the TeV $\gamma$-ray fluxes (eq. \ref{Apg}).
\section{Discussion and Results}
Considering that the one-zone homogeneous models are the most successful models to describe the SED of radio galaxies and taking into account that SSC models with only one electron population hardly explains the TeV $\gamma$-ray fluxes \citep{2005ApJ...634L..33G,2008A&A...478..111L},  we analyze the multifrequency data set of the radio galaxy IC310 in the context of homogeneous hadronic and leptonic models.  In the leptonic model, we take into account two electron populations to model the broadband SED whereas in the hadronic models, we consider that electromagnetic spectrum comes from  electrons and protons co-accelerated at the same emitting region of the jet. 
\subsection{Two electron populations}
The simplest leptonic model typically required to describe the emission from radio galaxy sources is the one-zone SSC. Within this picture, from radio to optical emission is produced by synchrotron radiation from electrons in a homogeneous, randomly-oriented magnetic field (B) and from X- to $\gamma$-rays are produced by inverse Compton scattering of the synchrotron photons by the same electrons which produce them.  However, the one-zone SSC model is not sufficient to explain the TeV emission from the core \citep{2005ApJ...634L..33G,2008A&A...478..111L}.  Therefore, we introduce a second electron population to describe the MAGIC emission and  a possible explanation for the  MAGIC observations is that the TeV emission is produced by another emitting region.   Using the method of Chi-square $ \chi^2$ minimization as implemented in the ROOT software package \citep{1997NIMPA.389...81B}, we fit the SED of IC310, as shown in Figure \ref{fig1}\footnote{The fitting process was integrated into a python script that called the ROOT routines via the pyroot module. The IC310 data were read from a file and written into an array which was then fitted with pyroot. The curves are smoothed using  the smooth sbezier function given in the gnuplot software (gnuplot.sourceforge.net)}.\\
From eqs. (\ref{synrad}) and (\ref{icrad}), we report in Table 1 the values obtained of  electron density, magnetic field, Doppler factor and size of emitting region, that describe the broadband SED of IC 310. In this fit, we require the effect of the extragalactic background light (EBL) absorption \citep{2008A&A...487..837F} and adopted the typical values reported in the literature such as the distance of IC310 ($d_z= 80\, {\rm Mpc}$) and the viewing angle ($\theta=16^\circ$) \citep{2015arXiv150201126S, 2014Sci...346.1080A, 2013ApJS..209...34A}.  Derived quantities such as the electron luminosities and the electron densities, the magnetic fields, etc, are also reported.  Table 1 shows all the parameter values obtained, used and derived in and from the fit.\\
\begin{center}\renewcommand{\arraystretch}{0.75}\addtolength{\tabcolsep}{2pt}
\begin{tabular}{ l c c c c c}
\hline
\hline
\normalsize{} & \normalsize{First population} &  \normalsize{Second population}&  \normalsize{Second population} \\
\normalsize{} & \normalsize{} &  \normalsize{Low state}&  \normalsize{High state} \\
\hline
\hline
\multicolumn{1}{c}{Obtained quantities} \\
\cline{1-1}
\scriptsize{$\delta_d$} & \scriptsize{1.8} &  \scriptsize{3.9}&  \scriptsize{4.9} \\
\scriptsize{$B$ (G)} & \scriptsize{0.12} &  \scriptsize{0.15}&  \scriptsize{0.125} \\
\scriptsize{$r_d$ (cm)} & \scriptsize{$3.98 \times 10^{16}\,$} & \scriptsize{$2.29 \times 10^{15}\,$}& \scriptsize{$2.38 \times 10^{15}\,$} \\
\scriptsize{$N_e$ (cm$^{-3}$)} & \scriptsize{$3.61\times 10^2$} &  \scriptsize{$1.58\times 10^4$}&  \scriptsize{$2.51\times 10^4$}\\
\scriptsize{$\alpha_e$} & \scriptsize{3.03} &  \scriptsize{3.14}&  \scriptsize{3.16} \\
 \\\hline
\multicolumn{1}{c}{Used quantities} & & & \\
\cline{1-1}
\scriptsize{$\gamma_{\rm e,min}$} & \scriptsize{$7.1\times10^2$} &  \scriptsize{$3.2\times10^5$}&  \scriptsize{$3.3\times10^5$} \\
\\  \hline
\multicolumn{1}{c}{Derived quatities} \\
\cline{1-1}
\scriptsize{$\gamma_{e,max}$} & \scriptsize{$1.34\times10^8$} &  \scriptsize{$1.20\times10^8$}&  \scriptsize{$1.31\times10^8$} \\
\scriptsize{$U_B\,\, {\rm (erg/cm^3)}$} & \scriptsize{$5.73\times 10^{-4}$} &  \scriptsize{$8.95\times 10^{-4}$}&  \scriptsize{$6.21\times 10^{-4}$}  \\
\scriptsize{$U_e\,\, {\rm (erg/cm^3})$} & \scriptsize{$1.75$} &  \scriptsize{$71.14$}&  \scriptsize{$109.07$}  \\
\scriptsize{$L_e\,\, {\rm (erg/s})$} & \scriptsize{$3.38\times 10^{44}$} &  \scriptsize{$1.02\times 10^{44}$}&  \scriptsize{$9.85\times 10^{44}$} \\
\hline
\hline
\end{tabular}
\end{center}
\begin{center}
\scriptsize{\textbf{Table 1. Parameters obtained, derived and used of leptonic model to fit  the spectrum of IC310. }}
\end{center}
As shown in Table 1, the second electron population that describes the TeV-$\gamma$-ray fluxes in high and low states is confined in one emitting region,  and the first electron population that explain the SED up to a few GeV is confined in an emitting region larger than that which confines to the second population.  From the different values obtained of emitting radius for the first and second electron population can be concluded that a multi-zone SSC emission is required to explain the broadband SED.  To obtain a good description of the TeV-$\gamma$-ray fluxes, the electron density required to model the high state must be higher than that used to describe the low  state.   Although the large value of the minimum Lorentz factor used for the first  population implies that electrons are efficiently accelerated by the Fermi mechanism, the value required for the second population is unrealistic.
\subsection{One electron and proton population}
If relativistic protons are presented in the jet of IC310, hadronic interactions must be considered for modeling the source emission.   Here, the ultra-relativistic electrons injected in the magnetized emitting region lose energy predominantly through synchrotron emission, that turn on serving as target photon field for inverse Compton scattering with the same electron population.   The instantaneously injected relativistic protons interact with the second SSC peak. The resulting GeV - TeV photons from pion decay products correspond to the energy band observed by the MAGIC telescopes.  Again, using the method of Chi-square $ \chi^2$ minimization as implemented in the ROOT software package \citep{1997NIMPA.389...81B}, we fit the SED of IC310, as shown in Figure \ref{fig2}\footnote{The fitting process was performed similarly to reported with the leptonic model}.  From both panels in Fig. \ref{fig2} can be seen that pion decay products coming from  photo-hadronic interactions can reproduce the MAGIC emission for high and low activity states without describing other data.
For these fits, we have considered again the effect of the EBL absorption \citep{2008A&A...487..837F} and adopted the typical values reported in the literature such as a the viewing angle ($\theta=16^\circ$),  the observed luminosities, the distance of IC310 ($d_z= 80\, {\rm Mpc}$), the minimum Lorentz factors and the energy normalizations  \citep{2015arXiv150201126S, 2014Sci...346.1080A, 2013ApJS..209...34A, 2014A&A...563A..91A,2013arXiv1308.0433E}.  Additionally, we have required similar values of power indexes for the injected relativistic electron and protons ($\alpha_e=\alpha_p$; 2011ApJ...736..131A). Table 2 shows all the parameter values obtained, used and derived in and from the fits.\\
As shown in Table 2,  the proton luminosity obtained with our model corresponds to a small fraction ($\sim 10^{-2}$) of the Eddington luminosity $L_{Edd}\sim 3.77\times 10^{46}$ erg/s, which was obtained taking into account the BH mass reported in IC310 ($3\times 10^8 M_\odot$; \citealp{2014Sci...346.1080A}).  By considering that acceleration $t'_{acc}\simeq\sqrt\frac\pi2m_p/q_e\, U^{-1/2}_B\,\gamma_p$ and cooling time scales for protons are similar,  the maximum proton Lorentz factor is {\small $ \gamma_{\rm p,max}=\left(\frac{9\, q_e^2}{8\pi\,\sigma_{T,p}^2}\right)^{1/4}\,U_B^{-1/4}$ where  $\sigma_{T,p}=\frac{m_e^2}{m_p^2}\,\sigma_{T}$. \cite{1978MNRAS.182..443B} discussed the particle distribution acceleration produced by a shock front.  Considering that particles need to be able to pass through the shock without  being strongly deflected, author found the resulting energy spectrum for protons and electrons. Although  this calculation was performed for acceleration of non-relativistic particles, can be seen that as the velocity of the shock increases, the ratio of proton and electrons densities decreases, tending to be similar (see table 1,  \citealp{1978MNRAS.182..443B}).  Due to the fact that the minimum Lorentz factor cannot be determined just assuming the standard scenario of injection and acceleration,  it is reasonable to use charge neutrality to justify a comparable number of electron and protons \citep{2013ApJ...768...54B,2009ApJ...704...38S, 2011ApJ...736..131A, 2014A&A...562A..12P}.   Requiring that electron and proton number densities are similar ($N_e\simeq N_p$), we have calculated that the minimum  proton Lorentz factors used to describe  the  TeV $\gamma$-ray fluxes in high and low states are $\gamma_{\rm p,min}=8\times 10^3$ and $5\times 10^4$, respectively. It is worth mentioning that the value of $\gamma_{\rm p,min}=1$ is excluded due to the proton luminosities become much higher than Eddington luminosity ($L_{\rm Edd}\ll L_{\rm p}$)}. The large value of $\gamma_{\rm e,min}$ provided by our model implies that electrons are efficiently accelerated by the Fermi mechanism only above this energy and that below this energy they are accelerated by a different mechanism that produces the hard electron distribution.\\
\begin{center}\renewcommand{\arraystretch}{0.75}\addtolength{\tabcolsep}{2pt}
\begin{tabular}{ l c c c c }
\hline
\hline
\normalsize{} & \normalsize{High state} &  \normalsize{Low state} \\
\hline
\hline
\multicolumn{1}{c}{Obtained quantities} \\
\cline{1-1}
\scriptsize{$\delta_d$} & \scriptsize{1.9} &  \scriptsize{1.8} \\
\scriptsize{$B$ (G)} & \scriptsize{1.3} &  \scriptsize{1.2} \\
\scriptsize{$r_d$ (cm)} & \scriptsize{$4.5 \times 10^{15}\,$} & \scriptsize{$5.0 \times 10^{15}\,$} \\
\scriptsize{$N_e$ (cm$^{-3}$)} & \scriptsize{$2.1\times 10^4$} &  \scriptsize{$1.9\times 10^4$} \\
\scriptsize{$\alpha_e$} & \scriptsize{3.05} &  \scriptsize{3.07} \\\hline

\multicolumn{1}{c}{Used quantities} & & \\
\cline{1-1}
\scriptsize{$\alpha_p$} & \scriptsize{3.05}&  \scriptsize{3.07} \\
\scriptsize{$\epsilon_0$}\,\,({\rm TeV}) & \scriptsize{1} &  \scriptsize{1}  \\
\scriptsize{$\gamma_{\rm e,min}$} & \scriptsize{$1\times10^3$} &  \scriptsize{$1\times10^3$} \\
\scriptsize{$L_{IC}\,\, {\rm (erg/s)}$} & \scriptsize{$1\times 10^{43}$} &  \scriptsize{$1\times 10^{43}$} \\\hline
\multicolumn{1}{c}{Derived quatities} \\
\cline{1-1}
\scriptsize{$\gamma_{\rm e,max}$} & \scriptsize{$4.25\times10^7$} &  \scriptsize{$4.25\times10^7$} \\
\scriptsize{$\gamma_{\rm p,min}$} & \scriptsize{$5.0\times10^4$} &  \scriptsize{$8.0\times10^3$} \\
\scriptsize{$\gamma_{\rm p,max}$} & \scriptsize{$2.8\times10^{10}$} &  \scriptsize{$2.8\times10^{10}$}  \\
\scriptsize{$\epsilon_{\rm peak}\,\, ({\rm MeV}) $ } & \scriptsize{$0.2$} &  \scriptsize{$0.2$}  \\
\scriptsize{$\epsilon_{\pi^0,\gamma,c}\,\,({\rm TeV})$} & \scriptsize{$0.52$} &  \scriptsize{$0.52$}  \\
\scriptsize{$U_B\,\, {\rm (erg/cm^3)}$} & \scriptsize{$5.7\times 10^{-2}$} &  \scriptsize{$5.7\times 10^{-2}$}  \\
\scriptsize{$U_e\,\, {\rm (erg/cm^3})$} & \scriptsize{$16.3$} &  \scriptsize{$16.3$}  \\
\scriptsize{$U_p\,\, {\rm (erg/cm^3})$} & \scriptsize{$17.1$} &  \scriptsize{$27.58$}  \\
\scriptsize{$L_p\,\, {\rm (erg/s)}$} & \scriptsize{$3.65\times 10^{44}$} &  \scriptsize{$4.3\times 10^{44}$}  \\
\scriptsize{$L_e\,\, {\rm (erg/s})$} & \scriptsize{$2.95\times 10^{44}$} &  \scriptsize{$2.95\times 10^{44}$} \\
\scriptsize{$E_{p,max}\,\, {\rm (EeV)}$} & \scriptsize{$24.1$} &  \scriptsize{$24.2$}  \\
\scriptsize{$N_{UHECRs}$} & \scriptsize{$0.28$} &  \scriptsize{$0.09$}  \\
\scriptsize{$N_{ev}$} & \scriptsize{$0.16$} &  \scriptsize{$0.02$}  \\
\hline
\hline
\end{tabular}
\end{center}
\begin{center}
\scriptsize{\textbf{Table 2. Parameters obtained, derived and used of lepton-hadronic model to fit  the spectrum of IC310.}}
\end{center}
After fitting the TeV $\gamma$-ray spectra of the radio galaxy IC310 with the neutral pion decay products, from eq. (\ref{numneu1}) we have obtained the neutrino fluxes and events expected from IC310.   Figure \ref{fig3} shows a sky-map with the 54 neutrino events reported by the IceCube collaboration, the 72 and 27 UHECRs collected by TA and PAO experiments, respectively. In addition, we have included a circular region of 5$^\circ$ around IC310.  As can be seen, there are neither neutrino track events nor UHECRs associated around IC310.\\
By assuming that the BH in the radio galaxy IC310 has the potential  to accelerate particles up to UHEs, the maximum energy achieved is 24.1 (24.2) EeV when the  TeV $\gamma$-ray flux in high (low) activity is considered. The previous result indicates that it is improbable that protons can be accelerated to energies as high as 57 EeV, although plausible for heavier accelerated ions.  From eq. \ref{Emax} can be seen that  any fluctuation in the strength of magnetic field and/or size of emitting region could confine protons above 57 EeV.  Interpreting the TeV gamma-ray spectra as p$\gamma$ interactions and extrapolating the accelerated proton spectrum up to energies higher than 1 EeV, the UHE proton spectra expected not only for IC310 but also for Cen A and M87 are plotted with the UHECR spectra collected with PAO \citep{2011arXiv1107.4809T}, HiRes \citep{2009APh....32...53H} and TA \citep{2013ApJ...768L...1A} experiment as shown in Figure \ref{fig4}. Left panel in this figure shows that the UHECR flux from the high state of IC 310 is lower (higher) than the Cen A (M87) and the right panel displays that the UHECR flux from the low state of IC 310 is the lowest flux.  In both panels are exhibited  that the contributions of the emitting regions to the UHECR fluxes are negligible.   Taking into account the exposure of TA experiment, we have estimated that the number of UHECRs above 57 EeV is 0.09 (0.28) when the TeV $\gamma$-ray flux in the low (high) state of activity is considered.  Due to the large scale of the Universe as well as strengths and geometries of magnetic fields: extragalactic and galactic \citep{2008ApJ...682...29D, 2010ApJ...710.1422R}, Galactic winds \citep{2016MNRAS.458..332H, 2016MNRAS.456.1723L, 2016yCat..51500081W}, magnetic winds \citep{1958ApJ...128..664P,2011ApJ...739...60E}, magnetic turbulence \citep{2013MNRAS.436.1245F, 2008Sci...320..909R}, UHECRs are deflected between the true direction to the source, and the observed arrival direction.  \cite{2010ApJ...710.1422R} estimated that the deflexion angle between the arrival direction of UHE protons and the sky position is quite large with a mean value of $<\theta_T>\simeq 15^\circ$. In addition, \cite{2008ApJ...682...29D} found that for observers situated within groups of galaxies like ours, about 70\% (35 \%) of UHECRs with energies above 60 EeV emerge inside $\sim 15^\circ$ ($5^\circ$), of the astrophysical object position.   We can see that number of UHECRs calculated with our model is consistent with those reported by the TA and/or PAO collaborations.  It is worth noting that the latter results reported by PAO suggested that UHECRs are heavy nuclei instead of protons \citep{2010PhRvL.104i1101A}.  If UHECRs have a heavy composition, then a significant fraction of nuclei must survive photodisintegrations in their sources. In this case, ultra-relativistic protons can be confined by the emitting regions so that these particles can achieve maximum energies of  $\gtrsim$ 54 EeV  for  Z$\gtrsim$ 2.\\ 
Interpreting the TeV $\gamma$-ray fluxes from IC310 as $\pi^0$ decay products from the p$\gamma$ interactions, then the number of neutrino events are 0.16  and 0.02 when these fluxes in high and low states are considered, respectively.   As neutrino fluxes were obtained from the p$\gamma$ interactions between accelerated protons and the seed photons at the SSC peaks, then fluctuations in the magnetic field, Doppler factor and/or emitting radius would change the seed photon density and finally the neutrino fluxes.  These fluxes for IC310,  Cen A and M87 were computed from relativistic protons interacting with the seed photons around the SSC peaks, as shown in Figure \ref{fig5}.  In this figure can be seen the neutrino spectra peaking around a few EeV, as expected from the interactions with seed photons around the synchrotron peak (the first SSC peak)  \citep{2008PhRvD..78b3007C}. Neutrino fluxes at TeV energies are obtained from the interactions with seed photons around the inverse Compton scattering  peak (the second SSC peak).   Considering the neutrino IceCube data \citep{2014PhRvL.113j1101A} and the upper limits at UHEs  set  by IceCube  (IC40; \cite{2011PhRvD..83i2003A}), PAO \citep{2011PhRvD..84l2005A}, RICE \citep{2012PhRvD..85f2004K} and ANITA \citep{2010PhRvD..82b2004G}  we confirm,  with our model,  the impossibility to observe nor TeV neither EeV neutrino events from IC310.  It is worth noting that if  heavy  nuclei would have been considered instead of protons,  HE neutrinos from UHE nuclei are significant lower than the neutrino flux obtained by protons. \\
\begin{center}\renewcommand{\arraystretch}{0.75}\addtolength{\tabcolsep}{2pt}
\begin{tabular}{ l c c c c c c}
\hline
\hline
\normalsize{} & \multicolumn{2}{c}{ \normalsize{IC310}}  & \normalsize{Cen A}& \normalsize{M87} \\
\normalsize{} & \normalsize{High state} &  \normalsize{Low state} \\
\hline
\hline
\scriptsize{$\delta_d$} & \scriptsize{1.9} &  \scriptsize{1.8} & \scriptsize{1.0} & \scriptsize{2.8}\\
\scriptsize{$B$ (G)} & \scriptsize{1.3} &  \scriptsize{1.2} &\scriptsize{3.6} & \scriptsize{1.61}\\
\scriptsize{$r_d$ (cm)} & \scriptsize{$4.5 \times 10^{15}\,$} & \scriptsize{$5.0 \times 10^{15}\,$}  & \scriptsize{$5.2 \times 10^{15}\,$}& \scriptsize{$2.1 \times 10^{15}\,$}\\
\scriptsize{$\alpha_p$} & \scriptsize{3.05}&  \scriptsize{3.07} &  \scriptsize{2.81} &  \scriptsize{2.80}\\
\scriptsize{$\gamma_{\rm p,min}$} & \scriptsize{$5.0\times10^4$} &  \scriptsize{$8.0\times10^3$} & \scriptsize{$1.0\times10^5$} &\scriptsize{$7.0\times10^6$}  \\
\scriptsize{$\gamma_{p,max}$} & \scriptsize{$2.8\times10^{10}$} &  \scriptsize{$2.8\times10^{10}$}  & \scriptsize{$6.73\times10^{10}$} & \scriptsize{$6.76\times10^{10}$}  \\
\scriptsize{$\epsilon_{\rm peak}\,\, ({\rm MeV}) $ } & \scriptsize{$0.2$} &  \scriptsize{$0.2$} &  \scriptsize{$0.1$} & \scriptsize{$0.5$} \\
\scriptsize{$\epsilon_{\pi^0,\gamma,c}\,\,({\rm TeV})$} & \scriptsize{$0.52$} &  \scriptsize{$0.52$} & \scriptsize{$0.32$}& \scriptsize{$0.54$} \\
\scriptsize{$U_B\,\, {\rm (erg/cm^3)}$} & \scriptsize{$5.7\times 10^{-2}$} &  \scriptsize{$5.7\times 10^{-2}$}  & \scriptsize{$0.52$}&\scriptsize{$0.10$} \\
\scriptsize{$U_p\,\, {\rm (erg/cm^3})$} & \scriptsize{$17.1$} &  \scriptsize{$27.58$}  &  \scriptsize{$2.55$} & \scriptsize{$2.69$} \\
\scriptsize{$L_p\,\, {\rm (erg/s)}$} & \scriptsize{$3.65\times 10^{44}$} &  \scriptsize{$4.3\times 10^{44}$}  & \scriptsize{$3.74\times 10^{43}$} & \scriptsize{$4.89\times 10^{43}$}\\
\scriptsize{$E_{p,max}\,\, {\rm (EeV)}$} & \scriptsize{$24.1$} &  \scriptsize{$24.2$} & \scriptsize{$40.1$} & \scriptsize{$6.55$}  \\
\scriptsize{$N_{UHECRs}$} & \scriptsize{$0.28$} &  \scriptsize{$0.09$} & \scriptsize{$1.52$} &  \scriptsize{$0.41$}\\
\hline
\hline
\end{tabular}
\end{center}
\begin{center}
\scriptsize{\textbf{Table 3.  Comparison of the values found  with the hadronic model used to describe the radio galaxies IC 310, Cen A and M87 .}}
\end{center}
In order to compare the parameters derived with our lepto-hadronic model for IC 310, Cen A and M87 \citep{2016ApJ...830...81F}, we summarize these values in Table 3. In this table can be noticed several features: i) although proton luminosities obtained in IC310 present the largest values, the numbers expected of UHECRs  are the smallest. It is  due to IC310 exhibits the softest spectral indexes, ii) as expected, the values of Doppler factors of IC 310 are in the range of Cen A and M87, which classify to IC310 as radio galaxy in stead of blazar and iii)  the values of the maximum energies that protons can achieve in the acceleration region indicate that only heavy ions can be expected from these radio galaxies.
\section{Summary and conclusions}
We have proposed leptonic and hadronic models to explain the broadband  SED observed in IC 310.  Taking into account that one electron population through the one-zone SSC model can explain the multiwavelength emission except for the non-simultaneous MAGIC emission, we have required additional electron and proton populations. In the leptonic framework, the additional electron population is confined in a different emitting region,  requiring a minimum Lorentz factor  exceedingly high $\gamma_e\sim 10^5$ which disfavors this model.   In the hadronic model, relativistic protons co-accelerated with the electrons that explain the SED up to a few GeV interact with seed photons at the second SSC peak.   The required proton luminosities are not extremely high $10^{44}$ erg/s, provided that the charge neutrality condition between the number of electrons and protons is given.\\
%
%
Correlating the TeV $\gamma$-ray and neutrino spectra through photo-hadronic interactions, we have estimated the fluxes and number of events expected in IceCube telescope, respectively.  We found that the neutrino flux produced by p$\gamma$ interactions  cannot explain the astrophysical flux and the expected $\nu_{\mu}$ events in the IceCube telescope are consistent with the nonneutrino track-like associated with the location of IC310 \citep{2013PhRvL.111b1103A, 2013Sci...342E...1I, 2014PhRvL.113j1101A}.\\
By considering that the proton spectrum given by a simple power law can be extended up to UHEs, we have  shown that in the radio galaxy IC310  neither from the emitting region nor other regions (in flaring intervals) UHECRs could be expected, which is in agreement with the TA and PAO observations.  Considering the flaring activities, cosmic rays would need a proton luminosity higher than the Eddington luminosity which is implausible.\\
On the other hand, If UHECRs have a heavy composition as suggested by PAO \citep{2010PhRvL.104i1101A}, UHE heavy nuclei in IC310  could be accelerated at energies below  $\sim$ 57 EeV, although the number of event expected would be less than one.   It is worth noting that if radio galaxies are the sources of UHECRs, their on-axis counterparts (i.e. blazars, and flat-spectrum radio quasars) should be considered a more powerful neutrino emitters \citep{2001PhRvL..87v1102A, 2014JHEAp...3...29D}.  In fact, a PeV neutrino shower-like was  recently associated with the flaring activity of the blazar  PKS B1424-418 \citep{2016NatPh..12..807K}. \\
%
%
%
%
Several  luminous blazars which are related with quasar-type AGN exhibit a double-humped shape characterized by a large luminosity ratio between  low and high energy humps.  \cite{2009ApJ...704...38S} studied those blazars with luminosity ratio between 10 - 100 which pose challenges to the standard SSC and hadronic models.  These authors concluded that hadronic models cannot account for the very hard X-ray spectra and also require extremely efficient acceleration of relativistic protons to the highest energies, surpassing the Eddington luminosity.\\
\cite{2013ApJ...768...54B} proposed leptonic and hadronic models to describe some of Fermi-LAT-detected blazars.  They found that either leptonic or hadronic models provide acceptable fit to the SEDs of most of the blazars, although in the hadronic case the proton luminosity demands values in the range $L_p\sim 10^{47} -10^{49}$ erg/s. Recently, \cite{2015MNRAS.450L..21Z} reviewed the hadronic model in radio-loud AGN, and the implications for the accretion in those sources. Authors found that the major problem in the hadronic models is related with highly super-Eddington jet power required to interpret the SEDs presented in AGNs. This problem, in turn would demand highly super-Eddington accretion rates and then, high luminosities which have not been observed. 
In our model, the proton luminosity is lower than the Eddington luminosity, similar to other hadronic models which have been successful in explaining, a PeV neutrino shower-like recently associated with the flaring activity of the blazar PKS B1424-418  \citep{2016NatPh..12..807K}, the SED of BL Lac Mrk421 in low state activity \citep{2011ApJ...736..131A} and radio galaxies such as Cen A and M87 \citep{2014ApJ...783...44F,2014MNRAS.441.1209F, 2015arXiv150407592K, 2016ApJ...830...81F, 2014A&A...562A..12P}.\\
The basic discrepancy between our model and those shown in  \cite{2009ApJ...704...38S} and \cite{2013ApJ...768...54B} lies in the distinct radiative processes used to describe the high-energy $\gamma$-ray  peak in the SED.  Whereas \cite{2013ApJ...768...54B} fitted the high-energy $\gamma$-ray peak with a substantial contribution of proton synchrotron radiation, we have used SSC emission, which is a more efficient process than that generated by proton synchrotron.  As pointed out by \citet{2009ApJ...704...38S}, given the characteristic proton- and electron-related synchrotron energies $\epsilon^{syn}_{\gamma,c,p}(\gamma_p=\gamma_e)=m_e/m_p\epsilon^{syn}_{\gamma,c,e}$ and the relation between the corresponding cooling rates $\mid \dot {\gamma}_p\mid_{\rm syn} (\gamma_p=\gamma_e)=(m_e/m_p) \mid \dot {\gamma}_e\mid_{\rm syn}$, the resulting efficiency of synchrotron proton emission becomes very low in comparison with synchrotron electron, although this efficiency can be increased by considering a large strength of magnetic field in the emitting region \citep{2000NewA....5..377A}.  By means of eq. \ref{Ae}, we can estimate the contribution of proton synchrotron radiation to the SED and also compare it with the electron synchrotron radiation. In this case, the relation of proton- and electron-synchrotron proportionality constants is $\frac{A_{\rm \gamma,syn,p}}{A_{\rm \gamma,syn}} \simeq \frac{\sigma_{T,p}\,U_{B,p}N_p}{\sigma_{T}\,U_BN_e}\, $, where $\sigma_{T,p}=\frac{m_e^2}{m_p^2}\,\sigma_{T}$ and  $U_{B,p}$ is the magnetic field energy density associated with the proton-synchrotron radiation used to fit the broadband SED.  Here, the values of Doppler and  minimum Lorentz factors have been considered of the same order \citep{2011ApJ...736..131A}.   Taking into account the charge neutrality condition $N_p=N_e$, the contribution of proton synchrotron emission is analyzed in two cases:
\begin{enumerate}
\item In our model, relativistic protons will radiate by synchrotron emission proportional to the magnetic field obtained after fitting the SED up to GeV energies. In this case, $U_{B,p}=U_B$ and then the  contribution of proton synchrotron emission in our model is very low $\frac{A_{\rm \gamma,syn,p}}{A_{\rm \gamma,syn}}\sim 10^{-6} $. 
Since the SED has been described successfully up to a few GeV with one-zone SSC model, proton synchrotron emission is not taken into account and then the proton luminosity normalized by photo-hadronic interactions does not increase.  As reported in this work,  moderate proton luminosities are required ($\sim 10^{44}$ erg/s) and weakly accreting disks are expected for IC 310.
\item In other models, some authors have reported that the typical values of magnetic field found after fitting the high-energy $\gamma$-ray peak of the SED  with proton synchrotron emission lies in the range of B$_p$=(10 - 100) B \citep[e.g. see][]{2013ApJ...768...54B,2009ApJ...704...38S}, with B the magnetic field obtained when a leptonic model is used to describe the SED. In other words, the magnetic field obtained for  proton synchrotron radiation is 10 - 100 times larger than that obtained for electron synchrotron radiation.  In this case,  the  proton and electron-related synchrotron contribution becomes $\frac{A_{\rm \gamma,syn,p}}{A_{\rm \gamma,syn}}\sim (10^{-5} - 10^{-3})$.   Normalizing the proton luminosities through proton-synchrotron emission, the proton luminosities is increased $\sim$ (10$^3$ - 10$^5$) times more than the electron luminosity, achieving values of  $\sim 10^{47} - 10^{49}$ erg/s. In this case, a highly super-Eddington jet power is required to interpret the broadband SEDs, as discussed by \cite{2015MNRAS.450L..21Z}. 
\end{enumerate}
Based on the previous analysis, we can conclude that a mixed model, with a  high-energy $\gamma$-ray peak being leptonic is more realistic than a hadronic model with a substantial contribution of proton synchrotron emission to the high-energy peak. It is worth noting that this model could be successfully used for the general population of blazars and radio galaxies.
\acknowledgments
We thank the anonymous referee for a critical reading of the paper and valuable suggestions that helped to improve the quality of this work. We also thank Maria Petropoulou, William Lee and Fabio de Colle for useful discussions.  This work was supported by UNAM PAPIIT grant IA102917.
%
%

\begin{thebibliography}{}
\expandafter\ifx\csname natexlab\endcsname\relax\def\natexlab#1{#1}\fi

\bibitem[{{Aartsen} {et~al.}(2013){Aartsen}, {Abbasi}, {Abdou}, {Ackermann},
  {Adams}, {Aguilar}, {Ahlers}, {Altmann}, {Auffenberg}, {Bai}, \&
  et~al.}]{2013PhRvL.111b1103A}
{Aartsen}, M.~G., {Abbasi}, R., {Abdou}, Y., {et~al.} 2013, Physical Review
  Letters, 111, 021103

\bibitem[{{Aartsen} {et~al.}(2014){Aartsen}, {Ackermann}, {Adams}, {Aguilar},
  {Ahlers}, {Ahrens}, {Altmann}, {Anderson}, {Arguelles}, {Arlen}, \&
  et~al.}]{2014PhRvL.113j1101A}
{Aartsen}, M.~G., {Ackermann}, M., {Adams}, J., {et~al.} 2014, Physical Review
  Letters, 113, 101101

\bibitem[{{Abbasi} {et~al.}(2011){Abbasi}, {Abdou}, {Abu-Zayyad}, {Adams},
  {Aguilar}, {Ahlers}, {Andeen}, {Auffenberg}, {Bai}, {Baker}, \&
  et~al.}]{2011PhRvD..83i2003A}
{Abbasi}, R., {Abdou}, Y., {Abu-Zayyad}, T., {et~al.} 2011, \prd, 83, 092003

\bibitem[{{Abbasi} {et~al.}(2005){Abbasi}, {Abu-Zayyad}, {Archbold}, {Atkins},
  {Bellido}, {Belov}, {Belz}, {BenZvi}, {Bergman}, {Boyer}, {Burt}, {Cao},
  {Clay}, {Connolly}, {Dawson}, {Deng}, {Fedorova}, {Findlay}, {Finley},
  {Hanlon}, {Hughes}, {Huntemeyer}, {Jui}, {Kim}, {Kirn}, {Knapp}, {Loh},
  {Maetas}, {Martens}, {Martin}, {Manago}, {Mannel}, {Matthews}, {Matthews},
  {O'Neill}, {Perera}, {Reil}, {Riehle}, {Roberts}, {Sasaki}, {Seman},
  {Schnetzer}, {Simpson}, {Smith}, {Snow}, {Sokolsky}, {Song}, {Springer},
  {Stokes}, {Thomas}, {Thomas}, {Thomson}, {Westerhoff}, {Wiencke}, {Zech}, \&
  {High Resolution Fly's Eye Collaboration}}]{2005ApJ...622..910A}
{Abbasi}, R.~U., {Abu-Zayyad}, T., {Archbold}, G., {et~al.} 2005, \apj, 622,
  910

\bibitem[{{Abdo} \& et~al.(2010)}]{2010ApJ...719.1433A}
{Abdo}, A.~A., \& et~al. 2010, \apj, 719, 1433

\bibitem[{{Abdo} {et~al.}(2011){Abdo}, {Ackermann}, {Ajello}, {Baldini},
  {Ballet}, {Barbiellini}, {Bastieri}, {Bechtol}, {Bellazzini}, {Berenji}, \&
  et~al.}]{2011ApJ...736..131A}
{Abdo}, A.~A., {Ackermann}, M., {Ajello}, M., {et~al.} 2011, \apj, 736, 131

\bibitem[{{Abraham} {et~al.}(2010){Abraham}, {Abreu}, {Aglietta}, {Ahn},
  {Allard}, {Allekotte}, {Allen}, {Alvarez-Mu{\~n}iz}, {Ambrosio},
  {Anchordoqui}, \& et~al.}]{2010PhRvL.104i1101A}
{Abraham}, J., {Abreu}, P., {Aglietta}, M., {et~al.} 2010, Physical Review
  Letters, 104, 091101

\bibitem[{{Abreu} {et~al.}(2011){Abreu}, {Aglietta}, {Ahn}, {Albuquerque},
  {Allard}, {Allekotte}, {Allen}, {Allison}, {Alvarez Castillo},
  {Alvarez-Mu{\~n}iz}, \& et~al.}]{2011PhRvD..84l2005A}
{Abreu}, P., {Aglietta}, M., {Ahn}, E.~J., {et~al.} 2011, \prd, 84, 122005

\bibitem[{{Abu-Zayyad} \& et~al.(2012)}]{2012NIMPA.689...87A}
{Abu-Zayyad}, T., \& et~al. 2012, Nuclear Instruments and Methods in Physics
  Research A, 689, 87

\bibitem[{{Abu-Zayyad} {et~al.}(2013){Abu-Zayyad}, {Aida}, {Allen}, {Anderson},
  {Azuma}, {Barcikowski}, {Belz}, {Bergman}, {Blake}, {Cady}, {Cheon}, {Chiba},
  {Chikawa}, {Cho}, {Cho}, {Fujii}, {Fujii}, {Fukuda}, {Fukushima}, {Hanlon},
  {Hayashi}, {Hayashi}, {Hayashida}, {Hibino}, {Hiyama}, {Honda}, {Iguchi},
  {Ikeda}, {Ikuta}, {Inoue}, {Ishii}, {Ishimori}, {Ivanov}, {Iwamoto}, {Jui},
  {Kadota}, {Kakimoto}, {Kalashev}, {Kanbe}, {Kasahara}, {Kawai}, {Kawakami},
  {Kawana}, {Kido}, {Kim}, {Kim}, {Kim}, {Kim}, {Kitamoto}, {Kitamura},
  {Kitamura}, {Kobayashi}, {Kobayashi}, {Kondo}, {Kuramoto}, {Kuzmin}, {Kwon},
  {Lan}, {Lim}, {Machida}, {Martens}, {Matsuda}, {Matsuura}, {Matsuyama},
  {Matthews}, {Minamino}, {Miyata}, {Murano}, {Myers}, {Nagasawa}, {Nagataki},
  {Nakamura}, {Nam}, {Nonaka}, {Ogio}, {Ohnishi}, {Ohoka}, {Oki}, {Oku},
  {Okuda}, {Oshima}, {Ozawa}, {Park}, {Pshirkov}, {Rodriguez}, {Roh},
  {Rubtsov}, {Ryu}, {Sagawa}, {Sakurai}, {Sampson}, {Scott}, {Shah}, {Shibata},
  {Shibata}, {Shimodaira}, {Shin}, {Shin}, {Shirahama}, {Smith}, {Sokolsky},
  {Stokes}, {Stratton}, {Stroman}, {Suzuki}, {Takahashi}, {Takeda}, {Taketa},
  {Takita}, {Tameda}, {Tanaka}, {Tanaka}, {Tanaka}, {Thomas}, {Thomson},
  {Tinyakov}, {Tkachev}, {Tokuno}, {Tomida}, {Troitsky}, {Tsunesada},
  {Tsutsumi}, {Tsuyuguchi}, {Uchihori}, {Udo}, {Ukai}, {Vasiloff}, {Wada},
  {Wong}, {Wood}, {Yamakawa}, {Yamane}, {Yamaoka}, {Yamazaki}, {Yang},
  {Yoneda}, {Yoshida}, {Yoshii}, {Zhou}, {Zollinger}, \&
  {Zundel}}]{2013ApJ...768L...1A}
{Abu-Zayyad}, T., {Aida}, R., {Allen}, M., {et~al.} 2013, \apjl, 768, L1

\bibitem[{{Ackermann} {et~al.}(2013){Ackermann}, {Ajello}, {Allafort},
  {Atwood}, {Baldini}, {Ballet}, {Barbiellini}, {Bastieri}, {Bechtol},
  {Belfiore}, {Bellazzini}, {Bernieri}, {Bissaldi}, {Bloom}, {Bonamente},
  {Brandt}, {Bregeon}, {Brigida}, {Bruel}, {Buehler}, {Burnett}, {Buson},
  {Caliandro}, {Cameron}, {Campana}, {Caraveo}, {Casandjian}, {Cavazzuti},
  {Cecchi}, {Charles}, {Chaves}, {Chekhtman}, {Cheung}, {Chiang}, {Chiaro},
  {Ciprini}, {Claus}, {Cohen-Tanugi}, {Cominsky}, {Conrad}, {Cutini},
  {D'Ammando}, {de Angelis}, {de Palma}, {Dermer}, {Desiante}, {Digel}, {Di
  Venere}, {Drell}, {Drlica-Wagner}, {Favuzzi}, {Fegan}, {Ferrara}, {Focke},
  {Fortin}, {Franckowiak}, {Funk}, {Fusco}, {Gargano}, {Gasparrini}, {Gehrels},
  {Germani}, {Giglietto}, {Giommi}, {Giordano}, {Giroletti}, {Godfrey},
  {Gomez-Vargas}, {Grenier}, {Guiriec}, {Hadasch}, {Hanabata}, {Harding},
  {Hayashida}, {Hays}, {Hewitt}, {Hill}, {Horan}, {Hughes}, {Jogler},
  {J{\'o}hannesson}, {Johnson}, {Johnson}, {Johnson}, {Kamae}, {Kataoka},
  {Kawano}, {Kn{\"o}dlseder}, {Kuss}, {Lande}, {Larsson}, {Latronico},
  {Lemoine-Goumard}, {Longo}, {Loparco}, {Lott}, {Lovellette}, {Lubrano},
  {Massaro}, {Mayer}, {Mazziotta}, {McEnery}, {Mehault}, {Michelson}, {Mizuno},
  {Moiseev}, {Monzani}, {Morselli}, {Moskalenko}, {Murgia}, {Nemmen}, {Nuss},
  {Ohsugi}, {Okumura}, {Orienti}, {Ormes}, {Paneque}, {Perkins},
  {Pesce-Rollins}, {Piron}, {Pivato}, {Porter}, {Rain{\`o}}, {Razzano},
  {Reimer}, {Reimer}, {Reposeur}, {Ritz}, {Romani}, {Roth}, {Saz Parkinson},
  {Schulz}, {Sgr{\`o}}, {Siskind}, {Smith}, {Spandre}, {Spinelli}, {Stawarz},
  {Strong}, {Suson}, {Takahashi}, {Thayer}, {Thayer}, {Thompson}, {Tibaldo},
  {Tinivella}, {Torres}, {Tosti}, {Troja}, {Uchiyama}, {Usher},
  {Vandenbroucke}, {Vasileiou}, {Vianello}, {Vitale}, {Werner}, {Winer},
  {Wood}, \& {Wood}}]{2013ApJS..209...34A}
{Ackermann}, M., {Ajello}, M., {Allafort}, A., {et~al.} 2013, \apjs, 209, 34

\bibitem[{{Aharonian}(2000)}]{2000NewA....5..377A}
{Aharonian}, F.~A. 2000, \na, 5, 377

\bibitem[{{Aharonian}(2002)}]{2002MNRAS.332..215A}
---. 2002, \mnras, 332, 215

\bibitem[{{Ahlers} \& {Murase}(2014)}]{2014PhRvD..90b3010A}
{Ahlers}, M., \& {Murase}, K. 2014, \prd, 90, 023010

\bibitem[{{Aleksi{\'c}} \& et~al.(2012)}]{2012A&A...539L...2A}
{Aleksi{\'c}}, J., \& et~al. 2012, \aap, 539, L2

\bibitem[{{Aleksi{\'c}} {et~al.}(2014{\natexlab{a}}){Aleksi{\'c}}, {Ansoldi},
  {Antonelli}, {Antoranz}, {Babic}, {Bangale}, {Barrio}, {Gonz{\'a}lez},
  {Bednarek}, {Bernardini}, {Biasuzzi}, {Biland}, {Blanch}, {Bonnefoy},
  {Bonnoli}, {Borracci}, {Bretz}, {Carmona}, {Carosi}, {Colin}, {Colombo},
  {Contreras}, {Cortina}, {Covino}, {Da Vela}, {Dazzi}, {De Angelis}, {De
  Caneva}, {De Lotto}, {Wilhelmi}, {Mendez}, {Prester}, {Dorner}, {Doro},
  {Einecke}, {Eisenacher}, {Elsaesser}, {Fonseca}, {Font}, {Frantzen}, {Fruck},
  {Galindo}, {L{\'o}pez}, {Garczarczyk}, {Terrats}, {Gaug}, {Godinovi{\'c}},
  {Mu{\~n}oz}, {Gozzini}, {Hadasch}, {Hanabata}, {Hayashida}, {Herrera},
  {Hildebrand}, {Hose}, {Hrupec}, {Idec}, {Kadenius}, {Kellermann}, {Kodani},
  {Konno}, {Krause}, {Kubo}, {Kushida}, {La Barbera}, {Lelas}, {Lewandowska},
  {Lindfors}, {Lombardi}, {Longo}, {L{\'o}pez}, {L{\'o}pez-Coto},
  {L{\'o}pez-Oramas}, {Lorenz}, {Lozano}, {Makariev}, {Mallot}, {Maneva},
  {Mankuzhiyil}, {Mannheim}, {Maraschi}, {Marcote}, {Mariotti},
  {Mart{\'{\i}}nez}, {Mazin}, {Menzel}, {Miranda}, {Mirzoyan}, {Moralejo},
  {Munar-Adrover}, {Nakajima}, {Niedzwiecki}, {Nilsson}, {Nishijima}, {Noda},
  {Orito}, {Overkemping}, {Paiano}, {Palatiello}, {Paneque}, {Paoletti},
  {Paredes}, {Paredes-Fortuny}, {Persic}, {Poutanen}, {Moroni}, {Prandini},
  {Puljak}, {Reinthal}, {Rhode}, {Rib{\'o}}, {Rico}, {Garcia}, {R{\"u}gamer},
  {Saito}, {Saito}, {Satalecka}, {Scalzotto}, {Scapin}, {Schultz}, {Schweizer},
  {Shore}, {Sillanp{\"a}{\"a}}, {Sitarek}, {Snidaric}, {Sobczynska}, {Spanier},
  {Stamatescu}, {Stamerra}, {Steinbring}, {Storz}, {Strzys}, {Takalo},
  {Takami}, {Tavecchio}, {Temnikov}, {Terzi{\'c}}, {Tescaro}, {Teshima},
  {Thaele}, {Tibolla}, {Torres}, {Toyama}, {Treves}, {Uellenbeck}, {Vogler},
  {Zanin}, {Kadler}, {Schulz}, {Ros}, {Bach}, {Krau{\ss}}, \&
  {Wilms}}]{2014Sci...346.1080A}
{Aleksi{\'c}}, J., {Ansoldi}, S., {Antonelli}, L.~A., {et~al.}
  2014{\natexlab{a}}, Science, 346, 1080

\bibitem[{{Aleksi{\'c}} {et~al.}(2014{\natexlab{b}}){Aleksi{\'c}}, {Antonelli},
  {Antoranz}, {Babic}, {Barres de Almeida}, {Barrio}, {Becerra Gonz{\'a}lez},
  {Bednarek}, {Berger}, {Bernardini}, {Biland}, {Blanch}, {Bock}, {Boller},
  {Bonnefoy}, {Bonnoli}, {Borla Tridon}, , \& {MAGIC
  Collaboration}}]{2014A&A...563A..91A}
{Aleksi{\'c}}, J., {Antonelli}, L.~A., {Antoranz}, P., {et~al.}
  2014{\natexlab{b}}, \aap, 563, A91

\bibitem[{{Aleksi{\'c}}(2010)}]{2010ApJ...723L.207A}
{Aleksi{\'c}}, J. e.~a. 2010, \apjl, 723, L207

\bibitem[{{Atoyan} \& {Dermer}(2001)}]{2001PhRvL..87v1102A}
{Atoyan}, A., \& {Dermer}, C.~D. 2001, Physical Review Letters, 87, 221102

\bibitem[{{Atoyan} \& {Dermer}(2003)}]{2003ApJ...586...79A}
{Atoyan}, A.~M., \& {Dermer}, C.~D. 2003, \apj, 586, 79

\bibitem[{{Becker}(2008)}]{2008PhR...458..173B}
{Becker}, J.~K. 2008, \physrep, 458, 173

\bibitem[{{Bell}(1978)}]{1978MNRAS.182..443B}
{Bell}, A.~R. 1978, \mnras, 182, 443

\bibitem[{{Bernardi} {et~al.}(2002){Bernardi}, {Alonso}, {da Costa}, {Willmer},
  {Wegner}, {Pellegrini}, {Rit{\'e}}, \& {Maia}}]{2002AJ....123.2990B}
{Bernardi}, M., {Alonso}, M.~V., {da Costa}, L.~N., {et~al.} 2002, \aj, 123,
  2990

\bibitem[{{B{\"o}ttcher} {et~al.}(2013){B{\"o}ttcher}, {Reimer}, {Sweeney}, \&
  {Prakash}}]{2013ApJ...768...54B}
{B{\"o}ttcher}, M., {Reimer}, A., {Sweeney}, K., \& {Prakash}, A. 2013, \apj,
  768, 54

\bibitem[{{Brown} \& {Adams}(2011)}]{2011MNRAS.413.2785B}
{Brown}, A.~M., \& {Adams}, J. 2011, \mnras, 413, 2785

\bibitem[{{Brun} \& {Rademakers}(1997)}]{1997NIMPA.389...81B}
{Brun}, R., \& {Rademakers}, F. 1997, Nuclear Instruments and Methods in
  Physics Research A, 389, 81

\bibitem[{{Cuoco} \& {Hannestad}(2008)}]{2008PhRvD..78b3007C}
{Cuoco}, A., \& {Hannestad}, S. 2008, \prd, 78, 023007

\bibitem[{{Das} {et~al.}(2008){Das}, {Kang}, {Ryu}, \&
  {Cho}}]{2008ApJ...682...29D}
{Das}, S., {Kang}, H., {Ryu}, D., \& {Cho}, J. 2008, \apj, 682, 29

\bibitem[{{Dermer} {et~al.}(2014){Dermer}, {Murase}, \&
  {Inoue}}]{2014JHEAp...3...29D}
{Dermer}, C.~D., {Murase}, K., \& {Inoue}, Y. 2014, Journal of High Energy
  Astrophysics, 3, 29

\bibitem[{{Dermer} {et~al.}(2009){Dermer}, {Razzaque}, {Finke}, \&
  {Atoyan}}]{2009NJPh...11f5016D}
{Dermer}, C.~D., {Razzaque}, S., {Finke}, J.~D., \& {Atoyan}, A. 2009, New
  Journal of Physics, 11, 065016

\bibitem[{{Eisenacher} {et~al.}(2013){Eisenacher}, {Colin}, {Lombardi},
  {Sitarek}, {Zandanel}, {Prada}, {Linfors}, {Paneque}, {Els{\"a}sser},
  {Mannheim}, {for the MAGIC Collaboration}, {M{\"u}ller}, {for the Fermi-LAT
  Collaboration}, {Dauser}, {Krau{\ss}}, {Wilbert}, {Kadler}, {Wilms}, {Bach},
  {Ros}, {Hovatta}, {for the OVRO team}, {Savolainen}, \& {for the MOJAVE
  team}}]{2013arXiv1308.0433E}
{Eisenacher}, D., {Colin}, P., {Lombardi}, S., {et~al.} 2013, ArXiv e-prints,
  arXiv:1308.0433

\bibitem[{{Everett} \& {Zweibel}(2011)}]{2011ApJ...739...60E}
{Everett}, J.~E., \& {Zweibel}, E.~G. 2011, \apj, 739, 60

\bibitem[{{Federrath}(2013)}]{2013MNRAS.436.1245F}
{Federrath}, C. 2013, \mnras, 436, 1245

\bibitem[{{Feretti} {et~al.}(1998){Feretti}, {Giovannini}, {Klein}, {Mack},
  {Sijbring}, \& {Zech}}]{1998A&A...331..475F}
{Feretti}, L., {Giovannini}, G., {Klein}, U., {et~al.} 1998, \aap, 331, 475

\bibitem[{{Fraija}(2014{\natexlab{a}})}]{2014ApJ...783...44F}
{Fraija}, N. 2014{\natexlab{a}}, \apj, 783, 44

\bibitem[{{Fraija}(2014{\natexlab{b}})}]{2014MNRAS.441.1209F}
---. 2014{\natexlab{b}}, \mnras, 441, 1209

\bibitem[{{Fraija} {et~al.}(2012){Fraija}, {Gonz{\'a}lez}, {Perez}, \&
  {Marinelli}}]{2012ApJ...753...40F}
{Fraija}, N., {Gonz{\'a}lez}, M.~M., {Perez}, M., \& {Marinelli}, A. 2012,
  \apj, 753, 40

\bibitem[{{Fraija} \& {Marinelli}(2016)}]{2016ApJ...830...81F}
{Fraija}, N., \& {Marinelli}, A. 2016, \apj, 830, 81

\bibitem[{{Franceschini} {et~al.}(2008){Franceschini}, {Rodighiero}, \&
  {Vaccari}}]{2008A&A...487..837F}
{Franceschini}, A., {Rodighiero}, G., \& {Vaccari}, M. 2008, \aap, 487, 837

\bibitem[{{Gaggero} {et~al.}(2015){Gaggero}, {Grasso}, {Marinelli}, {Urbano},
  \& {Valli}}]{2015ApJ...815L..25G}
{Gaggero}, D., {Grasso}, D., {Marinelli}, A., {Urbano}, A., \& {Valli}, M.
  2015, \apjl, 815, L25

\bibitem[{{Georganopoulos} {et~al.}(2005){Georganopoulos}, {Perlman}, \&
  {Kazanas}}]{2005ApJ...634L..33G}
{Georganopoulos}, M., {Perlman}, E.~S., \& {Kazanas}, D. 2005, \apjl, 634, L33

\bibitem[{{Ginzburg} \& {Syrovatskii}(1963)}]{1963SvA.....7..357G}
{Ginzburg}, V.~L., \& {Syrovatskii}, S.~I. 1963, \sovast, 7, 357

\bibitem[{{Gorham} {et~al.}(2010){Gorham}, {Allison}, {Baughman}, {Beatty},
  {Belov}, {Besson}, {Bevan}, {Binns}, {Chen}, {Chen}, {Clem}, {Connolly},
  {Detrixhe}, {de Marco}, {Dowkontt}, {Duvernois}, {Grashorn}, {Hill},
  {Hoover}, {Huang}, {Israel}, {Javaid}, {Liewer}, {Matsuno}, {Mercurio},
  {Miki}, {Mottram}, {Nam}, {Nichol}, {Palladino}, {Romero-Wolf}, {Ruckman},
  {Saltzberg}, {Seckel}, {Varner}, {Vieregg}, {Wang}, \& {ANITA
  Collaboration}}]{2010PhRvD..82b2004G}
{Gorham}, P.~W., {Allison}, P., {Baughman}, B.~M., {et~al.} 2010, \prd, 82,
  022004

\bibitem[{{Halzen}(2007)}]{2007Ap&SS.309..407H}
{Halzen}, F. 2007, \apss, 309, 407

\bibitem[{{Heesen} {et~al.}(2016){Heesen}, {Dettmar}, {Krause}, {Beck}, \&
  {Stein}}]{2016MNRAS.458..332H}
{Heesen}, V., {Dettmar}, R.-J., {Krause}, M., {Beck}, R., \& {Stein}, Y. 2016,
  \mnras, 458, 332

\bibitem[{{High Resolution Fly'S Eye Collaboration} {et~al.}(2009){High
  Resolution Fly'S Eye Collaboration}, {Abbasi}, {Abu-Zayyad}, {Al-Seady},
  {Allen}, {Amann}, {Archbold}, {Belov}, {Belz}, {Bergman}, {Blake}, {Brusova},
  {Burt}, {Cannon}, {Cao}, {Deng}, {Fedorova}, {Findlay}, {Finley}, {Gray},
  {Hanlon}, {Hoffman}, {Holzscheiter}, {Hughes}, {H{\"u}ntemeyer}, {Ivanov},
  {Jones}, {Jui}, {Kim}, {Kirn}, {Loh}, {Maestas}, {Manago}, {Marek},
  {Martens}, {Matthews}, {Matthews}, {Moore}, {O'Neill}, {Painter}, {Perera},
  {Reil}, {Riehle}, {Roberts}, {Rodriguez}, {Sasaki}, {Schnetzer}, {Scott},
  {Sinnis}, {Smith}, {Snow}, {Sokolsky}, {Springer}, {Stokes}, {Stratton},
  {Thomas}, {Thomas}, {Thomson}, {Tupa}, {Wiencke}, {Zech}, {Zhang}, {Zhang},
  {Zhang}, \& {High Resolution Fly's Eye Collaboration}}]{2009APh....32...53H}
{High Resolution Fly'S Eye Collaboration}, {Abbasi}, R.~U., {Abu-Zayyad}, T.,
  {et~al.} 2009, Astroparticle Physics, 32, 53

\bibitem[{{Hillas}(1984)}]{1984ARA&A..22..425H}
{Hillas}, A.~M. 1984, \araa, 22, 425

\bibitem[{{IceCube Collaboration}(2013)}]{2013Sci...342E...1I}
{IceCube Collaboration}. 2013, Science, 342, arXiv:1311.5238

\bibitem[{{Jiang} {et~al.}(2010){Jiang}, {Hou}, {Han}, {Sun}, \&
  {Wang}}]{2010ApJ...719..459J}
{Jiang}, Y.-Y., {Hou}, L.~G., {Han}, J.~L., {Sun}, X.~H., \& {Wang}, W. 2010,
  \apj, 719, 459

\bibitem[{{Jokipii}(1987)}]{1987ApJ...313..842J}
{Jokipii}, J.~R. 1987, \apj, 313, 842

\bibitem[{{Kadler} {et~al.}(2016){Kadler}, {Krau{\ss}}, {Mannheim}, {Ojha},
  {M{\"u}ller}, {Schulz}, {Anton}, {Baumgartner}, {Beuchert}, {Buson},
  {Carpenter}, {Eberl}, {Edwards}, {Eisenacher Glawion}, {Els{\"a}sser},
  {Gehrels}, {Gr{\"a}fe}, {Hase}, {Horiuchi}, {James}, {Kappes}, {Kappes},
  {Katz}, {Kreikenbohm}, {Kreter}, {Kreykenbohm}, {Langejahn}, {Leiter},
  {Litzinger}, {Longo}, {Lovell}, {McEnery}, {Phillips}, {Pl{\"o}tz}, {Quick},
  {Ros}, {Stecker}, {Steinbring}, {Stevens}, {Thompson}, {Tr{\"u}stedt},
  {Tzioumis}, {Wilms}, \& {Zensus}}]{2016NatPh..12..807K}
{Kadler}, M., {Krau{\ss}}, F., {Mannheim}, K., {et~al.} 2016, Nature Physics,
  12, arXiv:1602.02012

\bibitem[{{Kardashev} {et~al.}(1962){Kardashev}, {Kuz'min}, \&
  {Syrovatskii}}]{1962SvA.....6..167K}
{Kardashev}, N.~S., {Kuz'min}, A.~D., \& {Syrovatskii}, S.~I. 1962, \sovast, 6,
  167

\bibitem[{{Khiali} \& {de Gouveia Dal Pino}(2016)}]{2016MNRAS.455..838K}
{Khiali}, B., \& {de Gouveia Dal Pino}, E.~M. 2016, \mnras, 455, 838

\bibitem[{{Khiali} {et~al.}(2015){Khiali}, {de Gouveia Dal Pino}, \&
  {Sol}}]{2015arXiv150407592K}
{Khiali}, B., {de Gouveia Dal Pino}, E.~M., \& {Sol}, H. 2015, ArXiv e-prints,
  arXiv:1504.07592

\bibitem[{{Kravchenko} {et~al.}(2012){Kravchenko}, {Hussain}, {Seckel},
  {Besson}, {Fensholt}, {Ralston}, {Taylor}, {Ratzlaff}, \&
  {Young}}]{2012PhRvD..85f2004K}
{Kravchenko}, I., {Hussain}, S., {Seckel}, D., {et~al.} 2012, \prd, 85, 062004

\bibitem[{{Lemoine} \& {Waxman}(2009)}]{2009JCAP...11..009L}
{Lemoine}, M., \& {Waxman}, E. 2009, \jcap, 11, 9

\bibitem[{{Lenain} {et~al.}(2008){Lenain}, {Boisson}, {Sol}, \&
  {Katarzy{\'n}ski}}]{2008A&A...478..111L}
{Lenain}, J.-P., {Boisson}, C., {Sol}, H., \& {Katarzy{\'n}ski}, K. 2008, \aap,
  478, 111

\bibitem[{{Li} {et~al.}(2016){Li}, {Beck}, {Dettmar}, {Heald}, {Irwin},
  {Johnson}, {Kepley}, {Krause}, {Murphy}, {Orlando}, {Rand}, {Strong},
  {Vargas}, {Walterbos}, {Wang}, \& {Wiegert}}]{2016MNRAS.456.1723L}
{Li}, J.-T., {Beck}, R., {Dettmar}, R.-J., {et~al.} 2016, \mnras, 456, 1723

\bibitem[{{Longair}(1994)}]{1994hea2.book.....L}
{Longair}, M.~S. 1994, {High energy astrophysics. Volume 2. Stars, the Galaxy
  and the interstellar medium.}

\bibitem[{{Lovelace}(1976)}]{1976Natur.262..649L}
{Lovelace}, R.~V.~E. 1976, \nat, 262, 649

\bibitem[{{Lovelace}(1987)}]{1987A&A...173..237L}
---. 1987, \aap, 173, 237

\bibitem[{{Lovelace} {et~al.}(2005){Lovelace}, {Gandhi}, \&
  {Romanova}}]{2005Ap&SS.298..115L}
{Lovelace}, R.~V.~E., {Gandhi}, P.~R., \& {Romanova}, M.~M. 2005, \apss, 298,
  115

\bibitem[{{Lovelace} \& {Kronberg}(2013)}]{2013MNRAS.430.2828L}
{Lovelace}, R.~V.~E., \& {Kronberg}, P.~P. 2013, \mnras, 430, 2828

\bibitem[{{Lovelace} {et~al.}(2002){Lovelace}, {Li}, {Koldoba}, {Ustyugova}, \&
  {Romanova}}]{2002ApJ...572..445L}
{Lovelace}, R.~V.~E., {Li}, H., {Koldoba}, A.~V., {Ustyugova}, G.~V., \&
  {Romanova}, M.~M. 2002, \apj, 572, 445

\bibitem[{{Lovelace} {et~al.}(1994){Lovelace}, {Romanova}, \&
  {Newman}}]{1994ApJ...437..136L}
{Lovelace}, R.~V.~E., {Romanova}, M.~M., \& {Newman}, W.~I. 1994, \apj, 437,
  136

\bibitem[{{Lovelace} {et~al.}(1987){Lovelace}, {Wang}, \&
  {Sulkanen}}]{1987ApJ...315..504L}
{Lovelace}, R.~V.~E., {Wang}, J.~C.~L., \& {Sulkanen}, M.~E. 1987, \apj, 315,
  504

\bibitem[{{Marinelli} {et~al.}(2014){Marinelli}, {Fraija}, \&
  {Patricelli}}]{2014arXiv1410.8549M}
{Marinelli}, A., {Fraija}, N., \& {Patricelli}, B. 2014, ArXiv e-prints,
  arXiv:1410.8549

\bibitem[{{M{\"u}cke} \& {Protheroe}(2001)}]{2001APh....15..121M}
{M{\"u}cke}, A., \& {Protheroe}, R.~J. 2001, Astroparticle Physics, 15, 121

\bibitem[{{Murase} {et~al.}(2012){Murase}, {Dermer}, {Takami}, \&
  {Migliori}}]{2012ApJ...749...63M}
{Murase}, K., {Dermer}, C.~D., {Takami}, H., \& {Migliori}, G. 2012, \apj, 749,
  63

\bibitem[{{Neronov} {et~al.}(2010){Neronov}, {Semikoz}, \&
  {Vovk}}]{2010A&A...519L...6N}
{Neronov}, A., {Semikoz}, D., \& {Vovk}, I. 2010, \aap, 519, L6

\bibitem[{{Parker}(1958)}]{1958ApJ...128..664P}
{Parker}, E.~N. 1958, \apj, 128, 664

\bibitem[{{Petropoulou} {et~al.}(2014){Petropoulou}, {Lefa}, {Dimitrakoudis},
  \& {Mastichiadis}}]{2014A&A...562A..12P}
{Petropoulou}, M., {Lefa}, E., {Dimitrakoudis}, S., \& {Mastichiadis}, A. 2014,
  \aap, 562, A12

\bibitem[{{Pierre Auger Collaboration} \& et~al.(2007)}]{2007Sci...318..938P}
{Pierre Auger Collaboration}, \& et~al. 2007, Science, 318, 938

\bibitem[{{Pierre Auger Collaboration} \& et~al.(2008)}]{2008APh....29..188P}
---. 2008, Astroparticle Physics, 29, 188

\bibitem[{{Razzaque} {et~al.}(2012){Razzaque}, {Dermer}, \&
  {Finke}}]{2012ApJ...745..196R}
{Razzaque}, S., {Dermer}, C.~D., \& {Finke}, J.~D. 2012, \apj, 745, 196

\bibitem[{{Reynoso} {et~al.}(2011){Reynoso}, {Medina}, \&
  {Romero}}]{2011A&A...531A..30R}
{Reynoso}, M.~M., {Medina}, M.~C., \& {Romero}, G.~E. 2011, \aap, 531, A30

\bibitem[{{Rybicki} \& {Lightman}(1986)}]{1986rpa..book.....R}
{Rybicki}, G.~B., \& {Lightman}, A.~P. 1986, {Radiative Processes in
  Astrophysics}

\bibitem[{{Ryu} {et~al.}(2010){Ryu}, {Das}, \& {Kang}}]{2010ApJ...710.1422R}
{Ryu}, D., {Das}, S., \& {Kang}, H. 2010, \apj, 710, 1422

\bibitem[{{Ryu} {et~al.}(2008){Ryu}, {Kang}, {Cho}, \&
  {Das}}]{2008Sci...320..909R}
{Ryu}, D., {Kang}, H., {Cho}, J., \& {Das}, S. 2008, Science, 320, 909

\bibitem[{{Sahu} {et~al.}(2012){Sahu}, {Zhang}, \&
  {Fraija}}]{2012PhRvD..85d3012S}
{Sahu}, S., {Zhang}, B., \& {Fraija}, N. 2012, \prd, 85, 043012

\bibitem[{{Sari} \& {Esin}(2001)}]{2001ApJ...548..787S}
{Sari}, R., \& {Esin}, A.~A. 2001, \apj, 548, 787

\bibitem[{{Shklovskii}(1963)}]{1963SvA.....6..465S}
{Shklovskii}, I.~S. 1963, \sovast, 6, 465

\bibitem[{{Sijbring} \& {de Bruyn}(1998)}]{1998A&A...331..901S}
{Sijbring}, D., \& {de Bruyn}, A.~G. 1998, \aap, 331, 901

\bibitem[{{Sikora} {et~al.}(2009){Sikora}, {Stawarz}, {Moderski}, {Nalewajko},
  \& {Madejski}}]{2009ApJ...704...38S}
{Sikora}, M., {Stawarz}, {\L}., {Moderski}, R., {Nalewajko}, K., \& {Madejski},
  G.~M. 2009, \apj, 704, 38

\bibitem[{{Sitarek} {et~al.}(2015){Sitarek}, {Eisenacher Glawion}, {Mannheim},
  {Colin}, {for the MAGIC Collaboration}, {Kadler}, {Schultz}, {Krau{\ss}},
  {Ros}, {Bach}, \& {Wilms}}]{2015arXiv150201126S}
{Sitarek}, J., {Eisenacher Glawion}, D., {Mannheim}, K., {et~al.} 2015, ArXiv
  e-prints, arXiv:1502.01126

\bibitem[{{Sommers}(2001)}]{2001APh....14..271S}
{Sommers}, P. 2001, Astroparticle Physics, 14, 271

\bibitem[{{Spergel} {et~al.}(2003){Spergel}, {Verde}, {Peiris}, {Komatsu},
  {Nolta}, {Bennett}, {Halpern}, {Hinshaw}, {Jarosik}, {Kogut}, {Limon},
  {Meyer}, {Page}, {Tucker}, {Weiland}, {Wollack}, \&
  {Wright}}]{2003ApJS..148..175S}
{Spergel}, D.~N., {Verde}, L., {Peiris}, H.~V., {et~al.} 2003, \apjs, 148, 175

\bibitem[{{Stanev}(1997)}]{1997ApJ...479..290S}
{Stanev}, T. 1997, \apj, 479, 290

\bibitem[{{Stecker}(1968)}]{1968PhRvL..21.1016S}
{Stecker}, F.~W. 1968, Physical Review Letters, 21, 1016

\bibitem[{{Tavecchio} {et~al.}(1998){Tavecchio}, {Maraschi}, \&
  {Ghisellini}}]{1998ApJ...509..608T}
{Tavecchio}, F., {Maraschi}, L., \& {Ghisellini}, G. 1998, \apj, 509, 608

\bibitem[{{The Pierre Auger Collaboration} {et~al.}(2011){The Pierre Auger
  Collaboration}, {Abreu}, {Aglietta}, {Ahn}, {Albuquerque}, {Allard},
  {Allekotte}, {Allen}, {Allison}, {Alvarez Castillo}, \&
  et~al.}]{2011arXiv1107.4809T}
{The Pierre Auger Collaboration}, {Abreu}, P., {Aglietta}, M., {et~al.} 2011,
  ArXiv e-prints, arXiv:1107.4809

\bibitem[{Waxman \& Bahcall(1997)}]{PhysRevLett.78.2292}
Waxman, E., \& Bahcall, J. 1997, Phys. Rev. Lett., 78, 2292

\bibitem[{{Wiegert} {et~al.}(2016){Wiegert}, {Irwin}, {Miskolczi}, {Schmidt},
  {Mora}, {Damas-Segovia}, {Stein}, {English}, {Rand}, {Santistevan},
  {Walterbos}, {Krause}, {Beck}, {Dettmar}, {Kepley}, {Wezgowiec}, {Wang},
  {Heald}, {Li}, {MacGregor}, {Johnson}, {Strong}, {Desouza}, \&
  {Porter}}]{2016yCat..51500081W}
{Wiegert}, T., {Irwin}, J., {Miskolczi}, A., {et~al.} 2016, VizieR Online Data
  Catalog, 515

\bibitem[{{Zdziarski} \& {B{\"o}ttcher}(2015)}]{2015MNRAS.450L..21Z}
{Zdziarski}, A.~A., \& {B{\"o}ttcher}, M. 2015, \mnras, 450, L21

\end{thebibliography}
%

%
\begin{figure}
\centering
\includegraphics[width=1.1\textwidth]{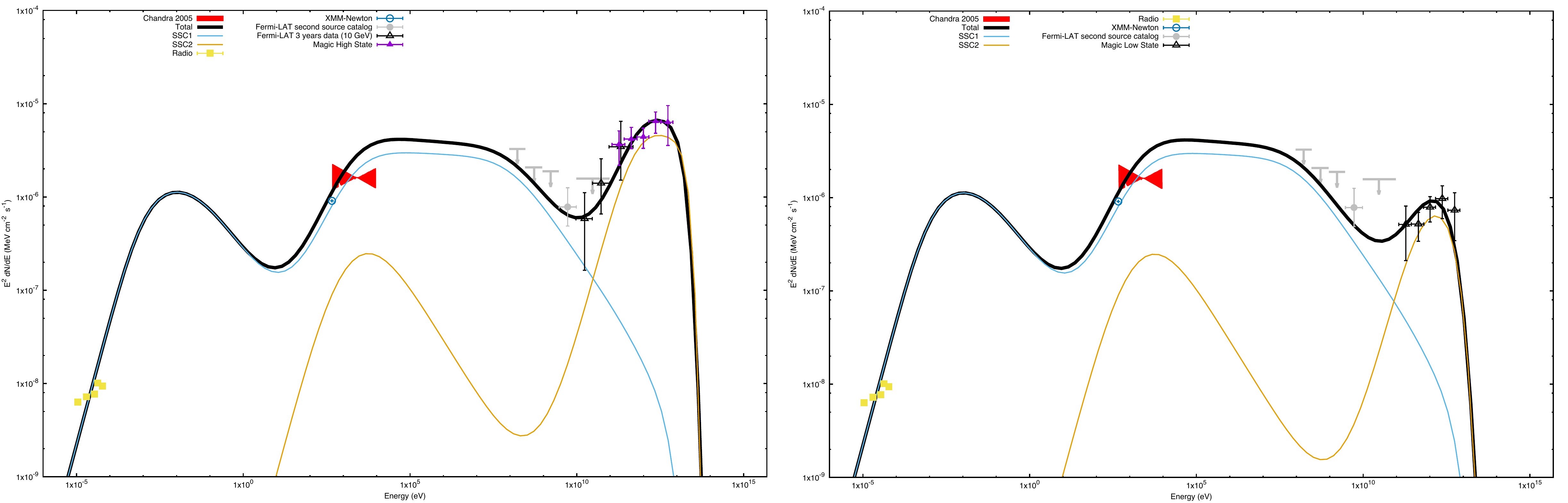}
\caption{SED of the IC310 described with a multi-zone SSC model given by two electron populations. The first electron population is used to model the SED up to a few GeV whereas the second one is required to fit the MAGIC observation.} \label{fig1}
\end{figure}
\begin{figure}
\centering
\includegraphics[width=1.1\textwidth]{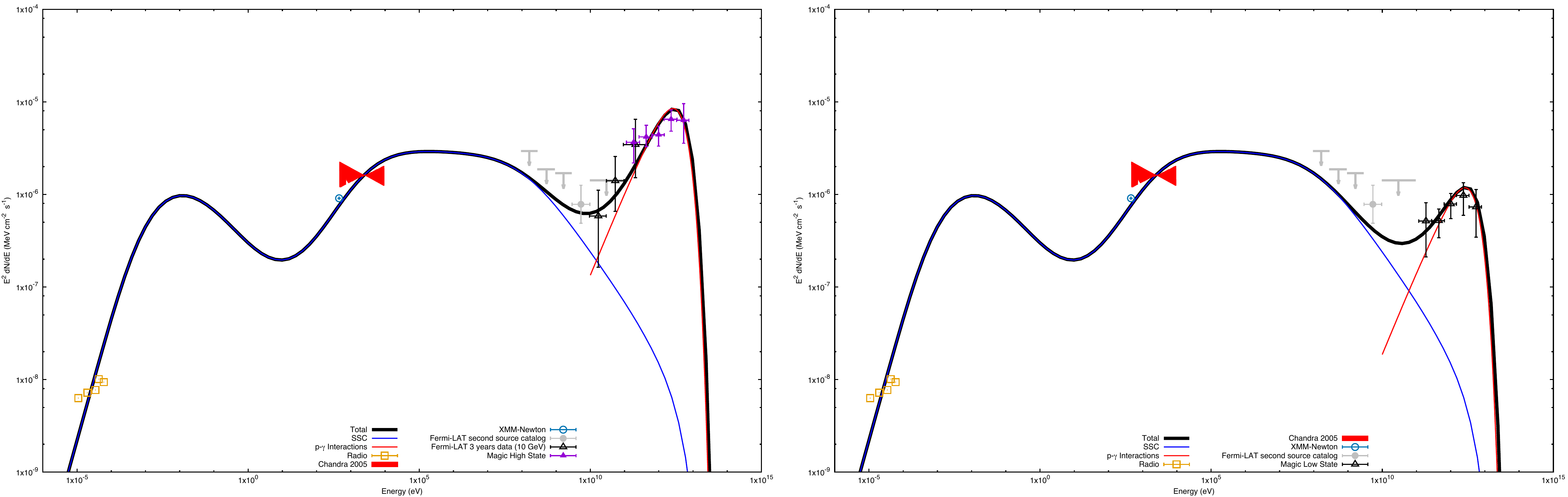}
\caption{SED of the IC310 described with a single-zone SSC model given by electron and proton populations. The electron population is required to fit the SED up to a few GeV and neutral pion decay products resulting from p$\gamma$ interactions is used to interpret the  TeV - GeV $\gamma$-ray spectra.} \label{fig2}
\end{figure}
\begin{figure}
\centering
\includegraphics[width=0.8\textwidth]{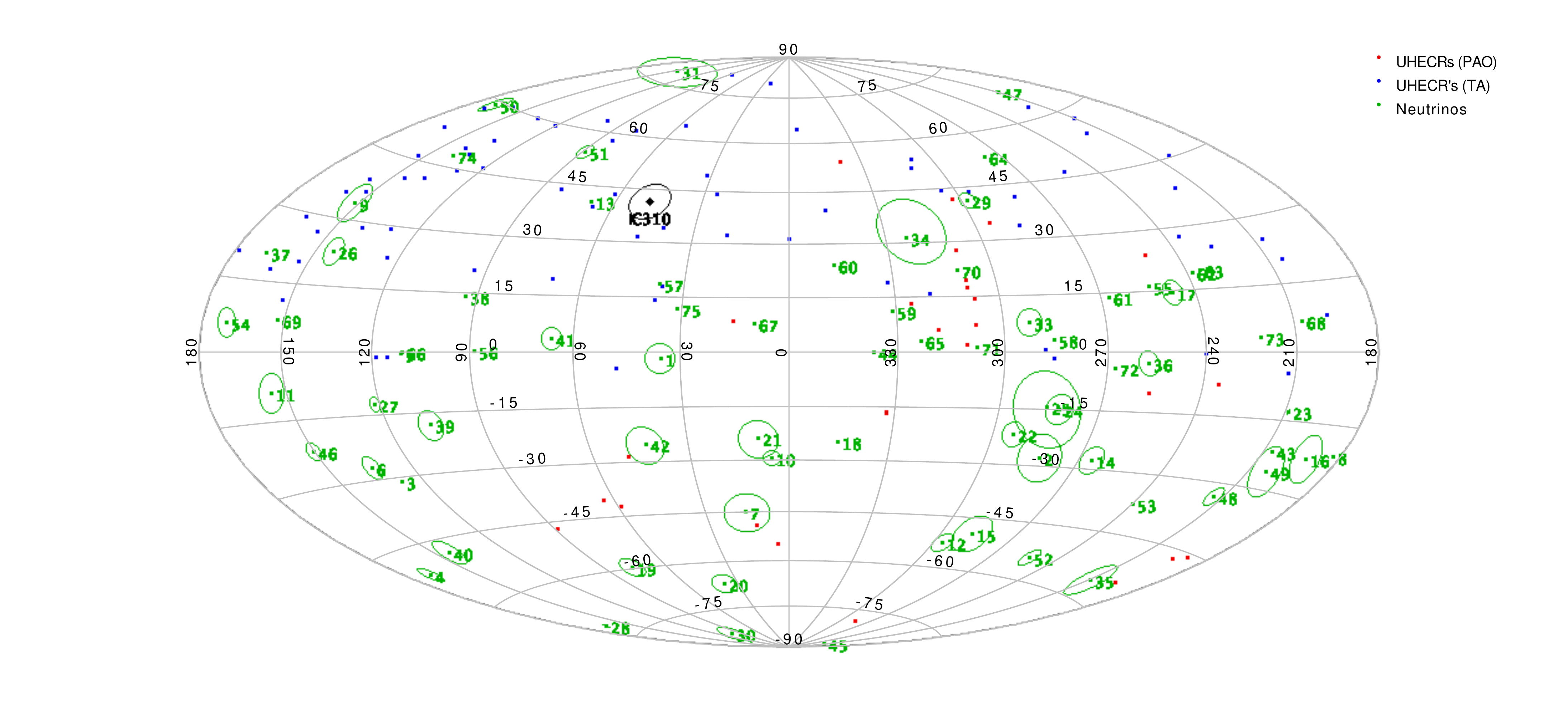}\\
\caption{Skymap in equatorial coordinates of UHECRs, neutrino events and IC310.  Blue and red points are the UHECRs reported by TA and PAO  collaborations, respectively. In green are the neutrino events reported by IceCube Collaboration.}\label{fig3}
\end{figure} 
\begin{figure}
\centering
\includegraphics[width=0.8\textwidth]{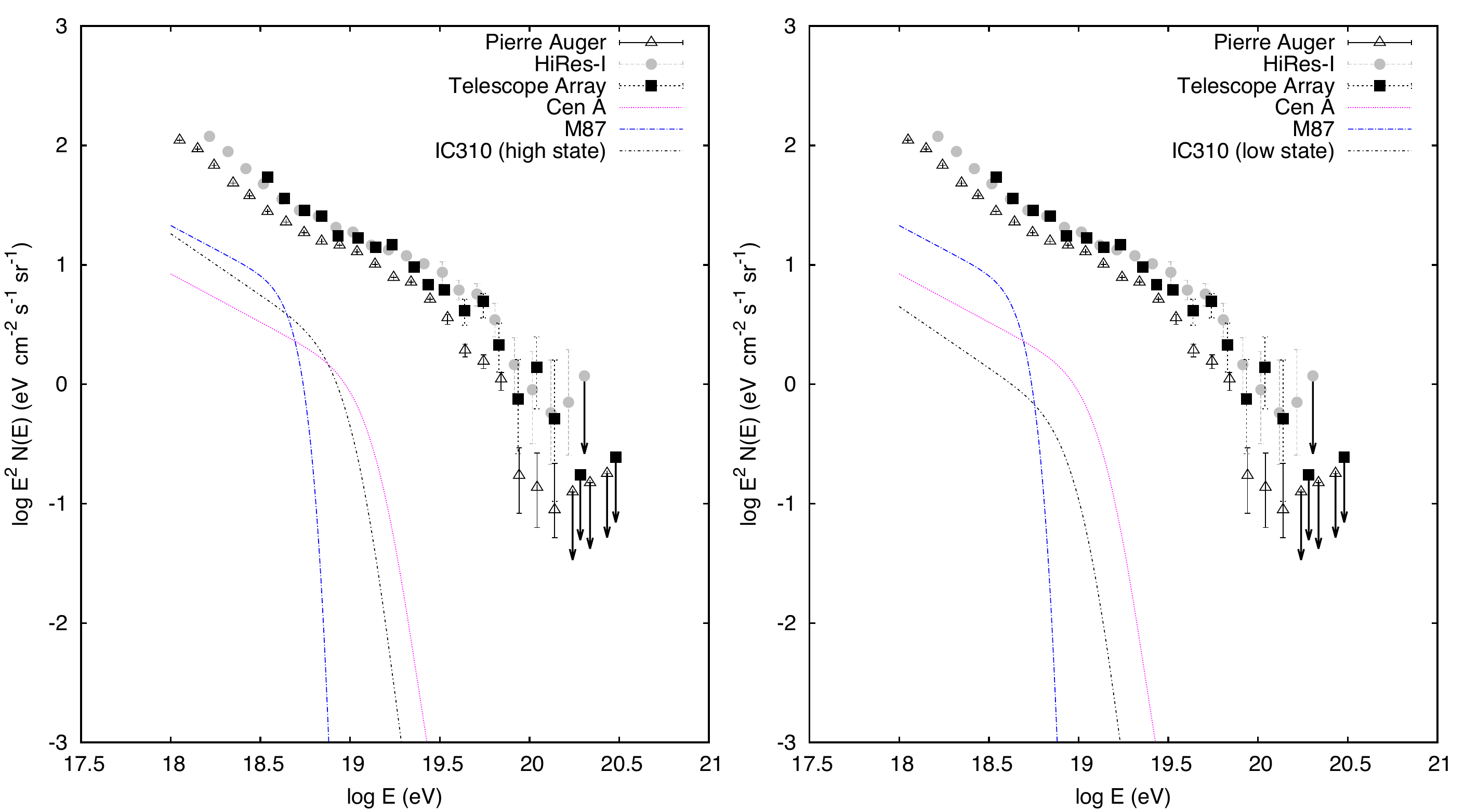}
\caption{The UHECR spectra collected with PAO \citep{2011arXiv1107.4809T}, HiRes \citep{2009APh....32...53H} and TA \citep{2013ApJ...768L...1A} experiment  are overlapped with UHE proton fluxes of IC310, Cen A and M87 resulting from extrapolating the proton fluxes used to describe the TeV $\gamma$-ray spectra.}. \label{fig4}
\end{figure}
\begin{figure}
\centering
\includegraphics[width=0.85\textwidth]{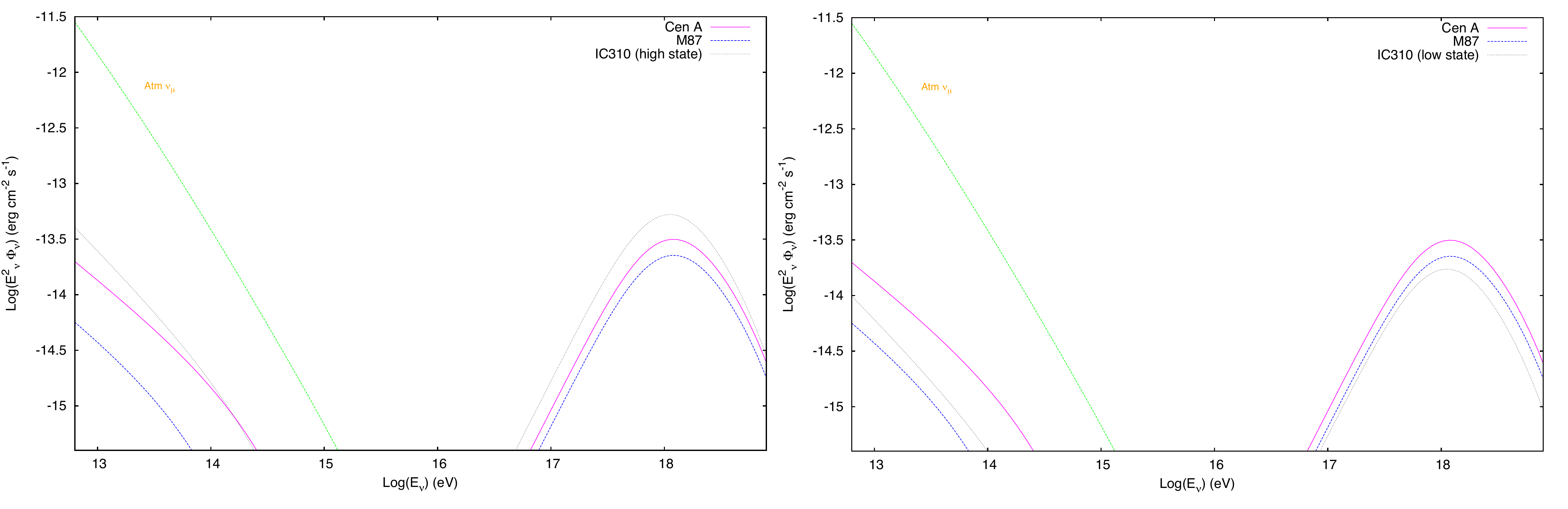}
\caption{Neutrino spectra of IC310, Cen A and M87 obtained as $\pi^\pm$ decay products from p$\gamma$ interactions of Fermi-accelerated protons with the seed photons around the both SED peaks.} \label{fig5}
\end{figure}
\end{document}